\PassOptionsToPackage{table,xcdraw,dvipsnames}{xcolor}
\documentclass[manuscript, screen]{acmart}
\usepackage{mathptmx}
\usepackage{amsmath}
\usepackage{csquotes}
\usepackage{balance}       % to better equalize the last page
\usepackage{graphics}      % for EPS, load graphicx instead 
\usepackage[T1]{fontenc}   % for umlauts and other diaeresis
\usepackage{color}
\usepackage{multirow}
\usepackage{subcaption}

\usepackage{paralist}

\DeclareCaptionLabelFormat{andtable}{#1~#2  \&  \tablename~\thetable}

\AtBeginDocument{%
  \providecommand\BibTeX{{%
    \normalfont B\kern-0.5em{\scshape i\kern-0.25em b}\kern-0.8em\TeX}}}

\setcopyright{none}
\acmDOI{}

\usepackage{microtype}        % Improved Tracking and Kerning
\usepackage{ccicons}          % Cite your images correctly!
\usepackage{graphicx}
\usepackage{subcaption}

\usepackage{todonotes}
\newcommand{\frameworkname}{AHA!}
\newcommand{\frameworknameCAP}{AHA!}
\newcommand{\para}[1]{\vspace{3pt}\noindent\textbf{#1}~}
\def\plaintitle{AHA!: Facilitating AI Impact Assessment by Generating Examples of Harms}

\def\emptyauthor{}
\def\plainkeywords{Responsible AI; impact assessment; LLM; crowdsourcing}

\definecolor{linkColor}{RGB}{6,125,233}
\hypersetup{%
  pdftitle={\plaintitle},
  pdfauthor={\emptyauthor},
  pdfkeywords={\plainkeywords},
  pdfdisplaydoctitle=true, % For Accessibility
  bookmarksnumbered,
  pdfstartview={FitH},
  colorlinks,
  citecolor=black,
  filecolor=black,
  linkcolor=black,
  urlcolor=linkColor,
  breaklinks=true,
  hypertexnames=false
}

\title{\plaintitle}

\author{Zana Bu\c{c}inca}
\authornote{Equal contribution}
\email{zbucinca@seas.harvard.edu}
\affiliation{%
  \institution{Harvard University}
  \city{Boston}
  \state{MA}
  \country{USA}
}

\author{Chau Minh Pham}
\authornotemark[1]
\email{ctpham@umass.edu}
\affiliation{%
  \institution{University of Massachusetts Amherst}
  \city{Amherst}
  \state{MA}
  \country{USA}
}

\author{Maurice Jakesch}
\email{mpj32@cornell.edu}
\affiliation{%
  \institution{Cornell University}
  \city{Ithaca}
  \state{NY}
  \country{USA}
}

\author{Marco Tulio Ribeiro}
\email{marco.correia@microsoft.com}
\affiliation{%
  \institution{Microsoft Research}
  \city{Redmond}
  \state{WA}
  \country{USA}
}

\author{Alexandra Olteanu}
\email{alexandra.olteanu@microsoft.com}
\affiliation{%
  \institution{Microsoft Research}
  \city{Montr\'eal}
  \state{Quebec}
  \country{Canada}
}

\author{Saleema Amershi}
\email{samershi@microsoft.com}
\affiliation{%
  \institution{Microsoft Research}
  \city{Redmond}
  \state{WA}
  \country{USA}
}
\begin{document}

\begin{abstract}
While demands for change and accountability for harmful AI consequences mount, foreseeing the downstream effects of deploying AI systems remains a challenging task. We developed \frameworkname{} (Anticipating Harms of AI), a generative framework to assist AI practitioners and decision-makers in anticipating potential harms and unintended consequences of AI systems prior to development or deployment.
Given an AI deployment scenario, \frameworkname{} generates descriptions of possible harms for different stakeholders.
To do so, \frameworkname{} systematically considers the interplay between common problematic AI behaviors as well as their potential impacts on different stakeholders, and narrates these conditions through vignettes. These vignettes are then filled in with descriptions of possible harms by prompting crowd workers and large language models.
By examining 4113 harms surfaced by \frameworkname{} for five different AI deployment scenarios, we found that \frameworkname{} generates meaningful examples of harms, with different problematic AI behaviors (e.g., false positives vs. false negatives) resulting in different types of harms. 
%Our results also demonstrate that the two generation sources -- crowds and the large language model -- complemented each other by generating together more diverse examples of harms to stakeholders than either crowd or the large language model alone.
Prompting both crowds (N = 98) and a large language model with the vignettes resulted in more diverse examples of harms than those generated by either the crowd or the model alone.  
To gauge \frameworkname{}'s potential practical utility, we also conducted semi-structured interviews with responsible AI professionals (N = 9). 
Participants found \frameworkname{}'s systematic approach to surfacing harms important for ethical reflection and discovered meaningful stakeholders and harms they believed they would not have thought of otherwise. Participants, however, differed in their opinions about whether \frameworkname{} should be used upfront or as a secondary-check and noted that AHA! may shift harm anticipation from an ideation problem to a potentially demanding review problem.
%Although they disagreed about where in the harm ideation process AHA! would be most helpful---raising concerns such as overreliance on AHA! and the need to cull through many examples---they found the systematic surfacing harms useful in supporting and structuring ethical reflection. %, informing responsible AI deployment decisions, thinking about mitigation strategies, and educating AI practitioners. 
Drawing on our results, we discuss design implications of building tools to help practitioners envision possible harms.
%Together, our results show how even seemingly small AI design and deployment decisions are value-laden and how practitioners need tools such as AHA!, that can help brainstorm about harms of their AI systems, surface the value tensions between stakeholders, and aid them in making more informed decisions that reflect their values.

%While all our participants found AHA! useful in supporting ethical reflection, they disagreed about where in the harm ideation process it would be most useful. Our participants also highlighted trade-offs [...]
%We conducted experiments and generated harms with AHA! for five different common scenarios of AI deployment. We evaluated the utility of our framework via semi-structured interviews with responsible AI experts and practitioners. Our results demonstrate that AHA! resulted in numerous AHA! moments --- with the framework participants discovered harms and stakeholders they had previously not thought about. They found AHA! valuable for ethical reflection and brainstorming about the adverse impacts of AI systems.[.....]
\end{abstract}

\maketitle
% ACM Classfication



% Author Keywords

% Print the classficiation codes
%\printccsdesc
%Please use the 2012 Classifiers and see this link to embed them in the text: %\url{https://dl.acm.org/ccs/ccs_flat.cfm}
%\includegraphics[trim= 0 100 0 145, clip,width=.92\linewidth]
\begin{figure}[t]
\includegraphics[trim= 0 2 0 10, clip,width=\linewidth]{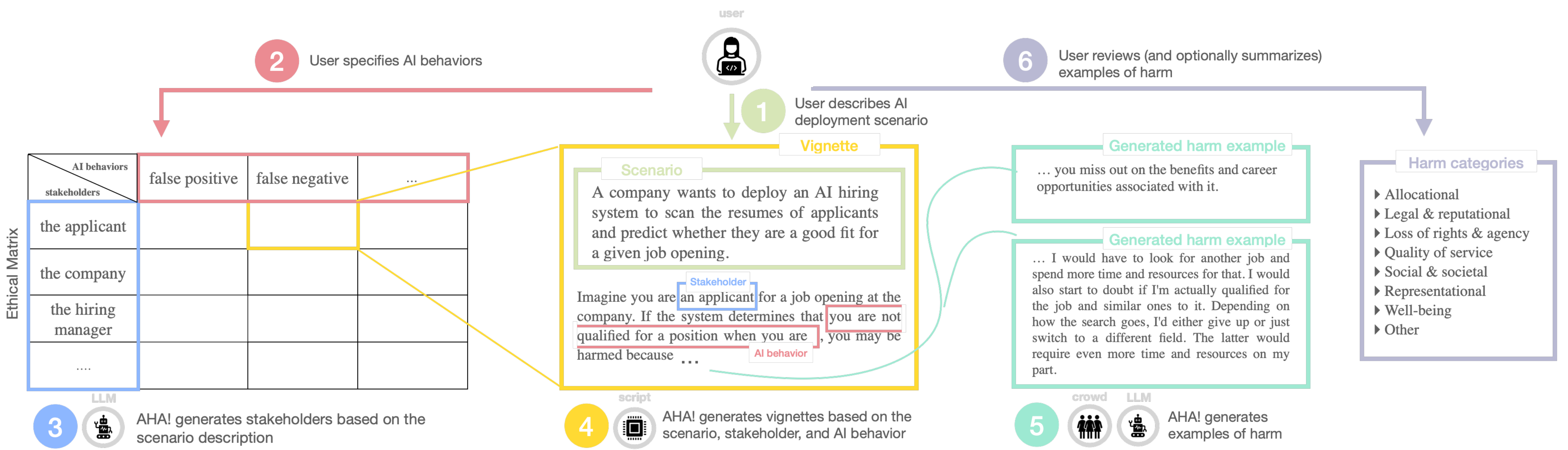}
%\vspace{-8pt}
\caption{Overview of \frameworkname{} for the hiring scenario. 
%1) Given a description of an AI deployment scenario, \frameworkname{} first generates relevant stakeholders. 2) Crossed with possible AI behaviors, these stakeholders form the Ethical Matrix, 3) for which \frameworkname{} then generates a separate vignette for each cell. 4) These vignettes along with the scenario description are then used to prompt crowd judges and an LLM to elicit examples of possible harms. 5) The examples can then be coded and clustered under high-level categories of harms to characterize them (optional).
1) Given a description of an AI deployment scenario and 2) a list of problematic AI behaviors of interests provided by a user, 3) \frameworkname{} first generates relevant stakeholders. Crossed with possible AI behaviors, these stakeholders form the Ethical Matrix, 4) for which \frameworkname{} then generates a separate vignette for each cell. 5) These vignettes along with the scenario description are then used to prompt crowd judges and an LLM to elicit examples of possible harms. 6) The examples can then be coded and clustered under high-level categories of harms to characterize them (optional).}

%Given a brief natural language description of a deployment scenario, \frameworkname{} generates descriptions of possible harms to different stakeholders that result from various problematic AI system behaviors. AI decision-makers can then review and reflect on the generated examples of harms before making decisions about whether and how to develop the AI system further.

%} 

\label{fig:main_fig}
%\vspace{-10pt}
\end{figure}

\section{Introduction}
In August 2020, thousands of UK students protested "F*ck the Algorithm" in response to the UK government employing an algorithm to predict their scores on A-level exams canceled due to the Covid-19 pandemic~\cite{kolkman2020f}. The algorithm disproportionately affected poorer students and led to 40\% of students receiving lower grades than expected on the exams that largely determine their university placement. While the backlash forced the government to retract the grades, the incident had already left students, families, and teachers scrambling to appeal the results and secure placement at universities~\cite{simonite2020skewed, lawless2020uk, adams2020alevel}, universities struggling to adjust resources due to unexpected admission numbers~\cite{fletcher2020half}, and government officials to contend with legal and reputational damage~\cite{coughlan2020sean}. The UK's automated grading fiasco is just one example of the growing number of incidents caused by AI systems deployed in applications and services sectors~\cite{mcgregor2021preventing}. %everyday applications and services as well as consequential industries like healthcare, justice, and employment sectors~\cite{mcgregor2021preventing}.

While demands for change and accountability for harmful AI consequences mount~\cite{eu}, foreseeing the downstream effects of deploying AI systems remains a challenging task~\cite{boyarskaya2020overcoming}. Anticipating harms requires grappling with the complexity and scope of contexts that even seemingly simple AI technologies may impact. The UK's automated grading system, for example, had a wide range of consequences that would have required envisioning the interplay between various stakeholders (from students and their families to universities and government agencies), problematic AI behaviors (e.g., inaccurate predictions, systematic biases), and how the AI might be used (e.g., admittance decisions, mitigation options, misuses). Even when guidance is given about how to proactively reflect on possible harms, the responsibility of carrying out this task most often falls on the shoulders of untrained and time-constrained practitioners such as system engineers or project managers~\cite{rakova2021responsible}. Moreover, even motivated practitioners can fail to envision downstream consequences to diverse stakeholders due to demographically skewed backgrounds and homogeneous experiences~\cite{crawford2016artificial, clark2016artificial, white2016preparing}.

In this paper, we present \frameworkname{} (Anticipating Harms of AI), a framework to support AI system creators, auditors and other decision-makers in foreseeing potential harms an AI system may cause before it is deployed or even implemented (see Figure~\ref{fig:main_fig} and Section~
\ref{sec:framework}). Given a short description of a deployment scenario, \frameworkname{} surfaces a set of potential harms by systematically considering how various stakeholders (e.g., individuals, groups, entities) experience problematic AI behaviors (e.g., false positives/negatives), narrating those experiences through vignettes (fictional scenarios used in the social sciences to elicit people's judgments when direct real-world investigations are impractical, unethical, or too complex to control~\cite{barter1999use, hidalgo2021humans}), and generating harms to complete the vignettes by prompting crowdworkers or large language models. The possible harms surfaced by \frameworkname{} are then reviewed by AI practitioners to inform development and deployment decisions. 

%[Need to mention here the objectives or benefits are a systematic/semi-automated approach to exploring the range of contexts and stakeholders, design to elicit empathy, and diversity of perspectives either from crowdworkers or LLMs (although I'm worried about saying LLMs provide diversity) ]

To evaluate \frameworkname{}'s capacity to surface possible harms, we first ran a series of experiments where we applied \frameworkname{} to five different AI deployment scenarios: hiring, loan application, content moderation, communication compliance, and disease diagnosis. Our analyses in Section~\ref{sec:quant} show that \frameworkname{} generates {\em meaningful} examples of possible harms (i.e., sensible and relevant to the given scenario). We also find that varying certain dimensions of problematic AI behaviors (e.g., false positives/negatives) significantly impacts the types of harms surfaced and that crowds and a large language model (GPT-3) surface comparable numbers of unique harms but significantly different {\em types of harms}. We further show that the combination of crowds + GPT3 produced significantly larger and more diverse examples of harms than either the crowd or the model alone. 
%Overall, \frameworkname{} surfaced different types of possible harms across scenarios, with a higher overlap among similar scenarios.

%In our experiments, we consider five different AI deployment scenarios (hiring, loan application, content moderation, communication compliance, and disease diagnosis). For each scenario, we generated examples of possible harms with \frameworkname{}---with both crowds (N=105) and GPT-3. 
%By examining the descriptions of possible harms generated by AHA! for five different AI deployment scenarios---with both crowds (N=105) and GPT-3, we found that our framework surfaces sensible examples of harms, with different problematic AI behaviors (e.g., false positives vs. false negatives) resulting in different types of harms. 
%To characterize the type of harms captured by our framework across scenarios, we qualitatively coded the surfaced examples to identify high-level types of harm. 
%\frameworkname{} also surfaced a different mix of possible harms across these scenarios, with a higher overlap among similar scenarios. 
%[insert other couple of findings from the analysis of generations; particularly about AI failures and stakeholders, which are called out in the earlier paragraph].
We then conducted a semi-structured interview study with (N = 9) Responsible AI professionals from industry and academia, where they reviewed the harms surfaced by our five experiments and gave us feedback on \frameworkname{}'s potential practical utility (see Section~\ref{sec:qual}). Our participants indicated that they found \frameworkname{}'s systematic approach to ethical reflection important for considering a broad range of harmful outcomes and discovered meaningful examples of possible harms they believed they would not have thought of on their own. However, they differed in opinion on whether practitioners should use \frameworkname{} upfront or as a secondary-check after engaging with the task of anticipating harms on their own. They also suggested that \frameworkname{} shifts harm anticipation from an ideation problem to a demanding review problem. Given these findings, we discuss implications and trade-offs in building tools for ethical reflection about AI systems (see Section~\ref{sec:discussion}).

\vspace{-2pt}
\section{Background and Related Work}

\noindent\textbf{Responsible AI Practices. }
%Calls for responsible AI are justifiably mounting alongside rising incidents of AI-related societal harms (e.g., [X]). This pressure has accelerated efforts to study [X], mitigate (e.g., [X]), and oversee or regulate (e.g., ~\cite{bernstein2021esr, raji2020closing}) risks associated with AI technologies. 
Realizing the promise of responsible AI \cite{microsoft,google,ibm,deloitte,montreal,eu} remains challenging in practice due to cultural, socio-technical, and organizational factors. Recommendations for addressing these challenges include ethical education~\cite{Oliver2021UndergraduateDS}, establishing ethical norms in AI communities ~\cite{liu2022examining, smith2022real}, aligning responsible AI goals with other organizational incentives~\cite{munn2022uselessness, rakova2021responsible}, introducing governance structures and review boards \cite{bernstein2021esr,raji2020closing,microsoftimpactassessment,leslie2019understanding}, and developing tools and processes to operationalize responsible AI goals~\cite{munn2022uselessness, rakova2021responsible, raji2020closing}. These are complementary approaches, with \frameworkname{} falling within the realm of tools to help development teams, auditors, and others anticipate AI risks and harms.

Within the scope of responsible AI tools and processes, several target specific stages in the AI development life cycle~\cite{suresh2021framework} including before~\cite{Ballard2019JudgmentCT, barnett2022crowdsourcing}, during~\cite{gebru2018datasheets, mitchell2018modelcards}, and after development~\cite{perez2022RedTeaming,ribeiro2022adatest}. Datasheets for datasets \cite{gebru2018datasheets} and model cards \cite{mitchell2018modelcards} aim to support documentation and reflection on decisions made during data collection and model training/testing. Others have proposed tools to support responsible AI decision making and documentation throughout product development~\cite{raji2020closing, madaio2020checklists}. \frameworkname{} is designed to help product teams proactively anticipate harms early and before resources have been spent making it increasingly difficult to change course~\cite{genus2018collingridge, rakova2021responsible}. Moreover, anticipating or identifying harms is often recommended as the first step in harm prevention or remediation~\cite{suresh2021framework,ONeil2020NearTermAI}, followed by activities such as cost-benefit analyses, root-cause analyses, and mitigation decision-making. \frameworkname{} may thus serve to support or connect responsible AI documentation and decision making at other stages of product development.    

\para{Anticipating Downstream Harms}
 of yet unbuilt or deployed technologies is fundamentally an exercise in creativity and imagination \cite{genus2018collingridge, boyarskaya2020overcoming}. Responsible AI guidance for anticipating harms often advises reflecting on the interplay between stated ethical principles or values, AI system behaviors, and intended contexts of use (e.g., who will be using the system and where and when), among other considerations~\cite{boyarskaya2020overcoming, ONeil2020NearTermAI}. Yet, each of these remain challenging to consider individually, let alone in combination. Many have criticized the vagueness and ambiguity of high-level AI ethical values and the subsequent difficulty practitioners face translating them into meaningful actions \cite{munn2022uselessness,liu2022examining, mittelstadt2019tooPrincipled, boyarskaya2020overcoming,jakesch2022different}, or have shown the difficulties practitioners have when foreseeing problematic system behaviors (e.g., failures or biases) or imagining how those behaviors might manifest in specific contexts~\cite{hong2021playbook,boyarskaya2020overcoming}. We ground harm anticipation in vignettes depicting \em problematic system behaviors \em manifesting in specific contexts of use and involving a range of \em stakeholders \em (see Section~\ref{sec:framework}). 
 %We demonstrate that through this grounding, \frameworkname{} can generate meaningful harms associated with violations of various ethical AI principles. % (e.g., fairness related harms of allocation and quality-of-service [X] as in Section X).   

Closest to our framework are approaches that propose systematic exploration of harmful outcomes by structuring the reflection process~\cite{mepham2006emtool,ONeil2020NearTermAI,smith2022real}. The Failure Modes and Effects Analysis (FMEA) approach~\cite{stadler2013fmea} prompts developers to enumerate potential system failures and then think through effects, possible causes, and mitigation actions. We structure the \frameworkname{} framework like an \em ethical matrix \em \cite{mepham2006emtool}, a conceptual tool originally developed to facilitate systematic consideration of biotechnology implications in food production \cite{mepham2000emFood}. We chose the ethical matrix, as it structures reflection around impacted stakeholders and ethical goals, concepts, or system behaviors, providing a more comprehensive view of AI impact than solely system-centered frameworks such as FMEA. 
O'Neil and Gunn \cite{ONeil2020NearTermAI} proposed the ethical matrix as a guide to systematically consider benefits and harms in the context of AI technologies. They structure the matrix by impacted stakeholders as rows and AI principles (e.g., justice), technical concepts (e.g., accuracy), and/or problematic outcomes (e.g., false positives/negatives) as columns. Ethical matrices are typically used to guide discussions amongst developers and decision-makers or amongst impacted stakeholders in participatory settings \cite{raji2020closing,johnson2021opening,lee2019webuildai}. Discussants are often invited to openly deliberate on issues concerning individual cells or participate in more guided workshops probing specific cell topics. Concerns, discussions, and evidence are then captured per cell for later reflection by decision-makers. 

Our work builds on these previous approaches for systematic exploration of harms in several important ways. First, we take a semi-automated approach to generate ethical matrices for given application scenarios, including automatically suggesting relevant stakeholders (rows) based on guidance about common stakeholder groups impacted by AI technologies~\cite{microsoftimpactassessment,ONeil2020NearTermAI} and automatically populating matrix cells with descriptive vignettes to help elicit empathy and ground reflection, a challenge emphasized by prior work \cite{mepham2006emtool, ONeil2020NearTermAI}. 
Second, we investigate the 
feasibility and effectiveness of using crowdsourcing or pre-trained language models to solicit or generate meaningful examples of harms.  
%effectiveness of diverse crowdworkers at generating meaningful harms to assess the potential of soliciting perspectives at lower-cost than typically required of participatory design approaches. 
Such solutions can encourage impact assessments of AI technologies beyond high-risk applications---as is typically recommended by most policies and frameworks due to their high-cost \cite{raji2020closing,leslie2019understanding,microsoftimpactassessment,raji2020closing}---by making conducting such assessments easier.

%Finally, we conduct experiments providing empirical evidence that our approach can surface meaningful harms for a variety of application scenarios (see Section X) and analyze the impact of various framework components including column manipulations (Section X) and generation sources (Section X) in surfacing distinct or diverse harms.  

%Jakesch et al.~\cite{jakesch2022different} developed a value elicitation survey to consult different demographic groups about their value priorities for AI systems in different contexts.

%- Harms can be caused by a variety of issues including adversarial uses, misuses, and unknown unknowns (e.g., unexpected behaviors caused by data biases or blind spots). In this work we focus on harms resulting from unintended consequences of common or expected but problematic AI behaviors (e.g., false positives/negatives) yet we expect our approach could extend to include these other sources.

%tarots of tech
%delphi method
%value sensitive design

\para{Envisioning Methods. }
Many methods have been proposed in the fields of cognitive psychology and human-computer interaction for spurring creative thinking and ideation about possible futures (for overviews see  \cite{lubart2001Creativity,nijstad2006ideas, frich2019creativity}). 
In the context of responsible AI, questions and prompts have been used to help AI developers envision potential harms~\cite{Ballard2019JudgmentCT,madaio2020checklists,microsoftimpactassessment}. Ballard et al ~\cite{Ballard2019JudgmentCT} designed a game to facilitate imagining harms through playing cards prompting players to craft hypothetical AI product reviews from various stakeholder perspectives and considering various ethical principles. Another approach to eliciting attitudes about hypothetical AI deployment futures %in the context of AI 
is the factorial survey~\cite{bailVignettes2019,janboecke2020adoptionVignettes}. Factorial surveys typically present participants with fictional narratives and solicit responses about those narratives, often manipulating dimensions of interest (e.g., contexts of use or system behaviors). These envisioning tools, particularly those leveraging groups or crowds \cite{nijstad2006ideas, rhys2021directed}, can help broaden perspectives of often homogenous research and development teams within the tech industry ~\cite{boyarskaya2020overcoming,ONeil2020NearTermAI,Ballard2019JudgmentCT}. %We leverage the factorial survey method by 
We draw on the factorial survey method and automatically generate partial vignettes for every cell in \frameworkname{}'s ethical matrix, and then use those vignettes as prompts to be completed with potential harms by crowds. 

Simulations are also widely used in computer science and other fields to help predict possible futures~\cite{agarwal2022multivariate, moore2022determinants}. Recent work has used large-language models (LLMs) to simulate and study human behavior. Horton~\cite{horton2022large} refers to LLMs as imperfect computational models of humans, or \em homo silicus\em, and used them to replicate past studies in behavioral economics. In another example, Park et al.~\cite{park2022social} used LLMs to simulate social interactions between distinct LLM-powered personas in social computing systems to help system designers refine community rules of engagement. 
%In this work, we also investigate the efficacy of LLMs for 
\frameworkname{} also uses LLMs to complete vignettes with possible examples of harms, complementing examples elicited from crowds.

\section{The \frameworknameCAP{} Framework Design}
\label{sec:framework}
\frameworkname{} is a generative framework designed to help AI developers, auditors, and other decision-makers anticipate harms prior to the development or deployment of AI systems. Figure~\ref{fig:main_fig} overviews the \frameworkname{} framework. Given a description of an AI deployment scenario, \frameworkname{} first generates potential \em stakeholders \em for that system via a large language model. Next, \frameworkname{} populates the cells of an \em ethical matrix\em, where the rows correspond to the generated \em stakeholders \em and the columns correspond to a pre-defined set of \em AI behaviors. \em
%\em common to classification-based systems. 
Each cell is populated with vignettes depicting scenarios where the corresponding stakeholder (row) experiences a behavior (column) of the AI system in question. Figure~\ref{fig:main_fig} shows a sample vignette for the \em hiring \em scenario. In this example, the stakeholder is an \em applicant \em experiencing a \em false negative \em behavior---the AI system determined they were not qualified for a position (incorrectly). Finally, \frameworkname{} prompts crowdworkers and a large language model to complete the vignettes with descriptions of harmful consequences. %The output is a set of such harms that decision-makers can review and reflect on. 
%

\iffalse
\begin{figure*}[t]
\includegraphics[width=\linewidth]{figures/system_overview.pdf}
\caption{System overview} \label{fig:system_overview}
\end{figure*}
\fi
\iffalse
\begin{figure}[t]
\includegraphics[trim= 0 9 0 7, clip,width=0.7\linewidth]{figures/example-vignette.png}
\vspace{-10pt}
\caption{Example vignette for the communication compliance scenario. For each \textit{scenario}, \frameworkname{} systematically manipulates both the \textit{stakeholder} and \textit{problematic AI behavior} to elicit completions about a diversity of possible \textit{harms}.} \label{fig:sample_vignette}
\vspace{-8pt}
\end{figure}
\fi
\para{Stakeholders (Rows). }
%Across domains (e.g., agriculture), decision-makers use a conceptual tool called \textit{The Ethical Matrix} to explore how interventions may affect different stakeholders~\cite{kaiser2007developing, oneilethicalmatrix}. Rows of such a matrix correspond to the stakeholders that may be affected by the intervention and columns may consist of the ethical principles they care about (e.g., \em justice, well-being\em). Because decision-making with an Ethical Matrix acknowledges and puts forth the value tension between different stakeholders, it is an invaluable tool to leverage for brainstorming about the impact of AI systems.
%
%As with any user-centered approach, 
Reasoning about the ethical implications of an AI system requires careful consideration of relevant stakeholders. 
Given the far-reaching consequences of many AI-based incidents~\cite{mcgregor2021preventing}, responsible AI guidance has started to emphasize consideration of a broad range of both direct and indirect stakeholders when envisioning possible harms~\cite{microsoftimpactassessment}. Direct stakeholders are considered those who directly interact with or are immediately affected by an AI system, while indirect stakeholders may be people associated with direct stakeholders or larger community groups. For example, in  a \em hiring \em scenario direct stakeholders may include the applicant and the hiring manager while indirect stakeholders may be the applicant's family, society at large, or future applicants. 
%In \frameworkname{}, these stakeholders make up the rows of our ethical matrix.

\para{AI Behaviors (Columns). }
Given that we are interested in harms arising in AI deployment scenarios, we chose to distinguish between various {\em problematic AI behaviors} that stakeholders may experience.
We make this distinction to set up fictitious scenarios---that stakeholders may find themselves in---in the vignettes we use as prompts to crowds and large language models. Problematic AI behaviors can correspond directly to model errors such as false positives and false negatives. However, not all errors are equal and factors such as how often they occur and how egregious they are can also lead to different outcomes.
We discuss ways in which the vignettes can operationalize these differences in Section~\ref{subsec:experimental_settings}.
%While we focus on \em problematic \em AI behaviors our framework supports the specification of any AI behavior --- including intended ones -- to elicit harms stemming from AI.

\para{Vignette Design (Matrix Cells). }
Given an AI use scenario, \frameworkname{} generates vignettes for each stakeholder (row) and problematic behavior (column) combination. We designed the vignettes to provide context and priming for e.g., crowdworkers and large language models to complete them with realistic descriptions of harms (see Figure~\ref{fig:main_fig} for an example vignette).\footnote{Since part of our goal was to understand how crowd and GPT-3 generations differ, to prevent any confounding, we prompted both with the same vignette design. Vignettes can be further tailored for the intended generation source.}
%\footnote{Pilot experiments were run in June-September 2021.}. 
%
Drawing on both psychology research showing that people view empathy as cognitively effortful and thus tend to avoid it when possible ~\cite{cameron2019empathy} and our early experimentation with different vignette designs,\footnote{We experimented with priming for both third-person (e.g., {\em imagine \underline{Joey} is a [stakeholder]}) and second-person (e.g., {\em imagine \underline{you} are a [stakeholder]}) perspectives, observing that crowdworkers seemed to engage more deeply when the vignettes were written in the second-person perspective.} we formulated the vignettes from a second-person perspective (e.g., {\em imagine \underline{you} are a [stakeholder]}). 
Furthermore, for scenarios where the use of the AI systems was ambiguous, we added more explicit language about how the systems' predictions will be leveraged. For example, in a communication compliance setting where ``{\em the AI system detected toxic language in an email of an employee}'', we specified that the system ``{\em will notify the employee's manager}''.\footnote{In our pilot experiments, leaving this unspecified resulted in (both crowdworkers and GPT-3) completing the vignettes with descriptions of possible problematic uses of the AI system rather than descriptions of possible harmful consequences.}

\vspace{-2pt}
\section{Characterizing the Harms Surfaced by \frameworkname{}}
\label{sec:quant}

To assess \frameworkname{}'s capacity to surface meaningful harms, we ran a series of experiments generating and then characterizing harms for five different hypothetical binary classification scenarios inspired by real-world AI incidents~\cite{mcgregor2021preventing}: hiring and loan applications, content moderation and communication compliance, and disease diagnosis.
Here, we describe our experiments and present findings from the exploratory data analysis.

\vspace{-2pt}
\subsection{Experiments}
\label{subsec:experimental_settings}
We applied \frameworkname{} as we envisioned practitioners would in practice. Specifically, after describing each AI scenario, we used \frameworkname{} to generate a list of relevant stakeholders and populate an ethical matrix with vignettes describing those stakeholders experiencing various problematic AI behaviors we were interested in. We then passed the vignettes to crowdworkers and a language model to complete. Finally, we manually reviewed and coded the harms for further analysis.

\para{Generating stakeholders.} 
To generate a range of both direct and indirect stakeholders for each scenario, we constructed a one-shot learning prompt for GPT-3. The prompt consisted of a context-specific list of stakeholders we manually defined for the \em hiring scenario \em based on consideration of relevant real-world incidents~\cite{mcgregor2021preventing}. It included: \em the applicant \em (for whom the AI system makes its hiring recommendation), \em other applicants \em and \em future applicants \em (whose job prospects may be impacted by the hiring recommendations for the target applicant or by use of the system), \em the hiring manager \em (who would rely on the AI system's recommendation), \em the tech company \em deploying the AI system and their \em HR team \em, \em the AI system developers \em (who may be accountable for the system's behavior), \em the applicant's family/friends, \em and applicants identifying with various demographic groups (e.g., racial, ethnic, or gender minorities and intersections).  
Given this prompt, \frameworkname{} then generated stakeholders for the other four scenarios, which we then refined before proceeding to the next stage.

\para{Selecting problematic AI behaviors. } Because we were interested in studying the impact of various problematic AI behaviors on what harms \frameworkname{} surfaces, we considered behaviors spanning several dimensions:

\begin{compactenum}[--]
\item \textit{False-positive / False-negative:} whether the system predicts an outcome when it should not or does not predict an outcome when it should. We distinguished between these errors because they often result in different real-world costs. In the hiring scenario, for example, if a system determines an applicant is qualified for a position when they are not, the company may be harmed by the applicant not being able to perform their duties. If the system determines an applicant is not qualified when they are, the tech company may miss an opportunity to hire an appropriate candidate. 
\item \textit{One-time / Accumulated:} whether the system makes a one-time error of that type or that error is made repeatedly or systematically over time. For example, a one-time false positive in the hiring scenario may go unnoticed but repeated hiring of unqualified candidates may reflect poorly on a hiring manager.
\item \textit{Egregious / Unspecified:} whether the system makes a severe error or an error of unspecified severity. Again in the hiring scenario, if the top candidate in a field is deemed unqualified for a tech company working in that field, the company may lose out on an opportunity to lead the field in that area.
\item \textit{Specified- / Unspecified-harm:} whether the vignette is conditioned on a specific harm (e.g., financial strain or emotional distress) or the harm is not-specified. For example, in the hiring scenario, the generated vignette might probe on \em financial strain \em by modifying the "you may be harmed because..." clause to "you may experience \textit{financial strain} because..." For these columns, practitioners using \frameworkname{} can specify harms they deem important for their scenario.
\end{compactenum}

%Given these four dimensions, 
\frameworkname{}'s ethical matrix thus included $2^4$ (sixteen) types of problematic behaviors (columns) per scenario. % in our experiment.

\begin{figure}[h]
\vspace{14pt}
\includegraphics[width=0.5\linewidth]{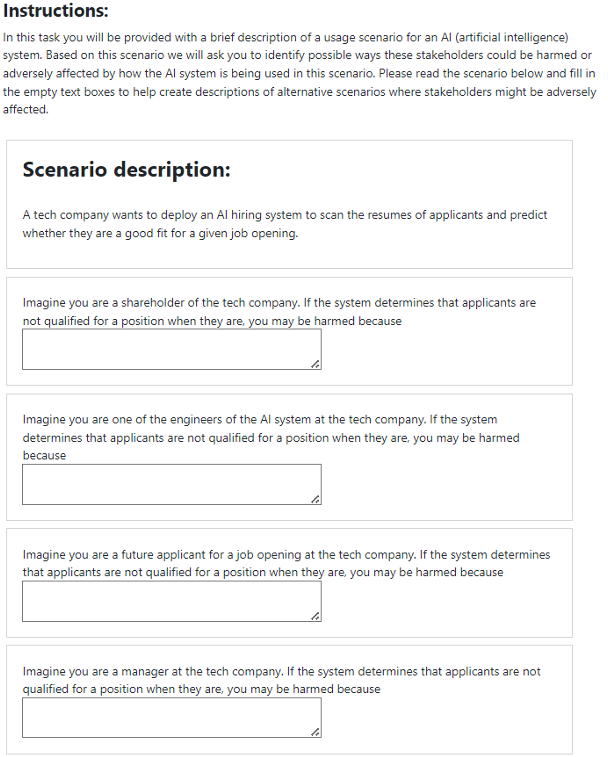}
\caption{A sample crowdsourcing task.} \label{fig:sample-task}
\end{figure}

\para{Completing vignettes with crowdsourcing. }
We recruited crowd judges from the Clickworker crowdsourcing platform (\url{www.clickworker.com}) in June-August 2022. Judges were first presented with a consent form before proceeding with the task. Each judge was then presented with four distinct vignettes from a single scenario, but for different stakeholders and problematic system behaviors.\footnote{We did so as in our pilot experiments, when prompted with multiple vignettes, the crowdworkers provided more diverse responses.} Vignettes were selected randomly from that scenario's corresponding ethical matrix until each vignette was filled out by three judges. After completing the task, judges were also prompted with demographic and background questions. Figure \ref{fig:sample-task} contains the example of a crowdsourcing task. Across all scenarios, we employed $105$ English speaking judges from North America. For each scenario, we capped the number of HITs per judge to 5. 

%via Microsoft's internal crowdsourcing tool---known as the Universal Human Relevance System (UHRS)~\cite{gray2016crowd,chen2019more}---that allows easy task design and quality control.  
%
To remove spammers and under-performing judges (and re-judge corresponding HITs), we used a mix of manual checks, speed checks, and attention checks. For manual checks, one of the authors read through the responses and flagged any gibberish (e.g. "abfbaufue") or absolutely irrelevant text (e.g. random concept definition copied directly from the Wikipedia). For speed checks, we determined those who submitted the task under five seconds to be spammers because this amount of time was too short to produce any meaningful open-ended answers. For attention checks, we incorporated a brief questions at the end of the task ("What is the color of the sky?"). We hypothesized that people who gave random answer like "green" were not paying full attention. On average, we flagged and requested re-judgments of $\approx40\%$ of responses to each task. 

Judges were paid for all HITs they performed, even when identified as spammers, paying on average about \$15 (USD) per hour.
In total, 98 trusted judges participated in our tasks.\footnote{The judges participating in our experiments span a range of self-reported ages (18-25: 17\%, 25-39: 49\%, 40-64: 31\%, 65+:2\%), familiarity with similar AI systems (with over 50\% of those reporting this saying to be slightly familiar or not familiar at all), and whether they believe to have experienced adverse impacts due to AI systems (with only ~15\% of those reporting this saying to have experienced adverse impacts from AI systems)} 
This crowdsourcing study was approved by our institution's IRB.
% We identified spammers through an attention check question and by manually checking their responses. We then remove and recollect generations for responses that look like obviously spams (i.e., keysmash) and fail the attention check question.

\para{Completing vignettes with a large language model. }
We also experimented with generating examples of harm for each of our five scenarios by prompting GPT-3 (davinci model with temperature set to $0.95$). For each scenario, we populated \frameworkname{}'s ethical matrix cells by prompting the model with the scenario description, the list of stakeholders, a few scenario-specific example harms for few-shot learning which we wrote by hand for random stakeholder-problematic AI behavior combinations (as we envision practitioners could do, e.g., for more obvious stakeholders), and the vignette to be completed. For each vignette (cell) we generated three different completions.\footnote{Through the iterative design of the prompts, we observed that more context about the task (i.e., providing the list of stakeholders coupled with the scenario description and the set of examples) led to more sensible completions from GPT-3. %We also observed that by increasing the temperature, a parameter that controls the randomness of the output, we obtained more generations that were not just paraphrased versions of our examples.
} 
%but were substantially different from the set of examples provided in the prompt. 

%[How did we prompt GPT-3? How many generations per cell? What was the context provided? What was the exclusion criteria?]

\para{Coding the examples of harm.}
We obtained a total of 4113 generations across all five scenarios. %\footnote{49.7\% of the generation stemmed from the crowd and 50.3\% from GPT-3}
To characterize the examples of harm generated by \frameworkname{}, we followed a thematic analysis approach~\cite{Braun2006UsingTA, Gibbs2007AnalyzingQD, nanayakkara2021unpacking}. First, one of the authors read the generations and open coded them to capture the types of harms produced. 
%The first annotator started by creating codes that best reflect the harms in each generation. 
%Some examples of these codes are harms that can describe or lead to a stakeholder "losing content or work due to the output of an AI system", or "experiencing repercussion or backlash from other people". 
Another author then reviewed the preliminary codes and iteratively aggregated them into broader harm taxonomy. Throughout the process, both annotators regularly compared annotations between different scenarios until categories stabilized. 

This process resulted in over 50 unique codes, grouped into 8 high-level categories:~\footnote{A single generation may have multiple codes and/or categories. In our analyses of distributions, we count a generation only once if it has multiple codes from the same high-level category.} 1) disparities in quality-of-service, 2) representational harms, 3) harms affecting people’s well-being, 4) legal and reputation harms, 5) allocational harms, 6) loss of rights or agency, 7) other social and societal harms, as well as 8) a catch-all category for harms not belonging in any other category. See Appendix~\ref{appendix:harms_taxonomy} for a detailed overview of the resulting harm taxonomy. We refer to the higher-level harms (e.g., allocational) as {\em harm categories} and lower-level codes (e.g., opportunity loss) as {\em harm subcategories}.

%We use this harm taxonomy to understand the harms that emerge from crowd and GPT-3 generations. We observed that, within a category, certain types of harms tend to be more prevalent than others. For instance, harms that describe or can lead to economic strain, losing money/customers/business, or dealing with raising costs were the most prevalent in the loan approval scenario. Similarly, for hiring, harms that describe or can lead to stakeholders missing out on an opportunity are most prevalent. This type of harm also appears in all scenarios. 

\begin{figure}[t]
    \begin{subfigure}[T]{0.62\textwidth}
    \includegraphics[trim= 20 5 10 3, clip,width=\linewidth]{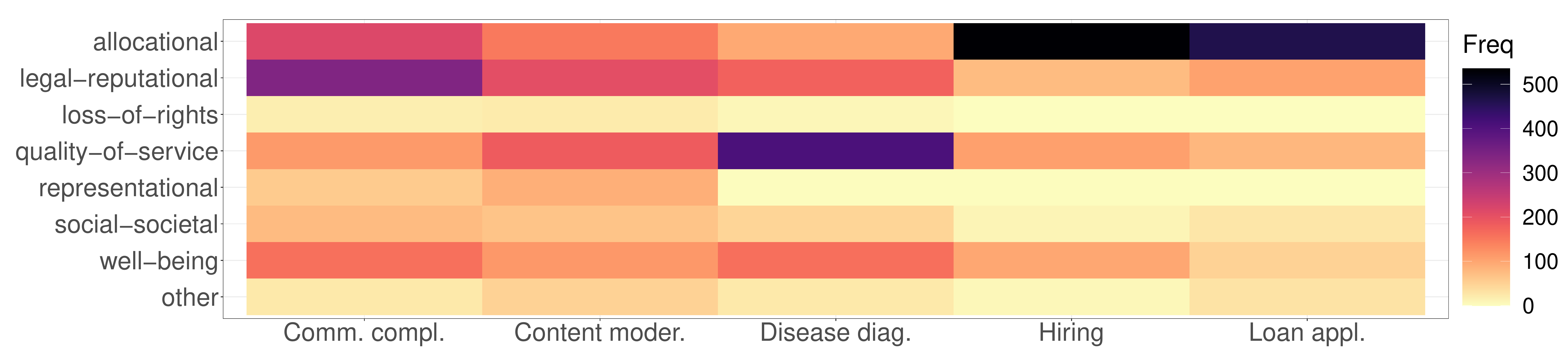}
    \end{subfigure}
    \quad
    \begin{subfigure}[T]{0.33\textwidth}
    \scriptsize
    \def\arraystretch{0.85}
    \begin{tabular}{@{}p{0.7cm}|p{0.75cm}p{0.75cm}p{0.7cm}p{0.5cm}}
    %\hline
     & \textbf{Comm. compl.} & \textbf{Content moder.} & \textbf{Disease diag.} & \textbf{Hiring} \\ \hline
    \rowcolor[HTML]{EFEFEF} 
    \textbf{Content moder.}         & {\color[HTML]{9B9B9B}.71 (n.s.)}   & -   & -    & -  \\
    \textbf{Disease\qquad diag.}    & {\color[HTML]{9B9B9B}.71 (n.s.)}   & {\color[HTML]{9B9B9B}.31 (n.s.)} & -    & -  \\
    \rowcolor[HTML]{EFEFEF} 
    \textbf{Hiring}                     & \color[HTML]{9A0000} \textless{}.0001   & {\color[HTML]{9A0000} \textless{}.0001}  & {\color[HTML]{9A0000} \textless{}.0001}  & -  \\
    \textbf{Loan\qquad applic.}          & {\color[HTML]{9A0000} \textless{}.0001}  & {\color[HTML]{9A0000} \textless{}.0001}  & {\color[HTML]{9A0000} \textless{}.0001}  & {\color[HTML]{CE6301} .02} \\ \hline
    \end{tabular}
    \end{subfigure}

    %\captionlistentry[table]{A table beside a figure}
    %\captionsetup{labelformat=andtable}
    \vspace{-14pt}
    \caption{Distribution of harm categories across scenarios. \em p-values \em for $\chi^2$ pairwise comparisons with Holm-Bonferroni corrections.}
    \label{fig:distribution-scenarios}
    \vspace{-4pt}
\end{figure}

% R1: Does AHA generate sensible examples of possible harms?
% R2: Does varying dimensions of problematic AI behaviors impact the harms generated?
% R3: Does eliciting harms from crowds vs GPT3 impact the harms generated?
% R4: Do experienced responsible AI professionals see value in AHA for helping teams anticipate harms in practice?

\subsection{RQ1: Does \frameworkname{} generate meaningful examples of possible harms?}
We consider a harm {\em meaningful} if it is not nonsense, is an actual harmful outcome, and is relevant to the scenario in question. To understand whether \frameworkname{} surfaces meaningful harms, we first look at the prevalence of generations coded as {\em not meaningful}. Only 7\% (288) of the examples elicited from crowds and GPT-3 were coded as {\em not meaningful} (63.2\% from crowds). Of these, 86.4\% were \em nonsensical \em (not clear or does not make sense in context) while the remaining were \em not a harm \em (sensible but does not represent an actual harm to a stakeholder).

To understand whether \frameworkname{}'s surfaces harms relevant to the given deployment scenarios, we look at whether the harms differ across scenarios in a meaningful way. That is, we would expect different scenarios to generate different harm categories (e.g., the \em hiring \em scenario could be considered quite different than the {\em content moderation} scenario in terms of stakes and context) and more similar scenarios to generate similar harm categories (e.g., the {\em hiring} and {\em loan application} scenarios are similar in that they are both instances of a resource allocation problem, while the {\em content moderation} and {\em communication compliance} scenarios are similar in that they are both instances of online content moderation policy enforcement). 

%These scenarios can be roughly grouped into three primary use cases: resource allocation (hiring and loan application), online moderation policy enforcement (content moderation and communication compliance), and diagnostics (disease diagnosis).

To test this hypothesis, we a ran Chi-square ($\chi^2$) analysis on the distribution of harm generated and grouped into harm categories, followed by post-hoc $\chi^2$ pairwise comparisons of scenarios with Holm-Bonferroni correction to account for multiple comparisons. We found significant differences in the distribution of harm categories across scenarios ($\chi^2$ (28, N = 4382) = 1577, p <.0001). Our results also show that the distribution of harm categories was more similar for similar scenarios (Figure~\ref{fig:distribution-scenarios}).
For example, both the \em hiring \em and \em loan application \em scenarios, where the AI system's role is to distribute benefits to decision-subjects, surfaced mainly {\em allocational} harms.\footnote{In our taxonomy, {\em allocational harms} are defined as: when the AI system's output affects the allocation of resources or opportunities relating to finance, education, employment, healthcare, housing, insurance, or social welfare.} 
Whereas, the \em communication compliance \em and \em content moderation scenarios, \em where the AI system's decision can result in punitive outcomes for decision-subjects, surfaced more \em legal and reputational \em harms. Finally, for the \em disease diagnosis \em scenario, where the AI's decision can impact one's well-being, the harms concentrated more around {\em quality of service}.\footnote{ In our taxonomy, {\em quality of service} harms are defined as: when either the AI system or a service the AI system is used for does not work equally well for different individuals or groups or doesn't work as intended.}

%\subsection{Scenarios: Harm Distributions Across Scenarios}
\begin{table}[t]
\centering
\scriptsize
\def\arraystretch{1}
\begin{tabular}{@{}lll|ll|ll|ll|ll|ll@{}}
%\toprule
\multicolumn{3}{c|}{\textbf{}}  & 
\multicolumn{2}{l|}{\cellcolor[HTML]{EFEFEF} False positive / negative}   & 
\multicolumn{2}{l|}{\cellcolor[HTML]{EFEFEF} Harm specified / not specified}  & 
\multicolumn{2}{l|}{\cellcolor[HTML]{EFEFEF} Accumulated / one-time}  & 
\multicolumn{2}{l|}{\cellcolor[HTML]{EFEFEF} Egregious / not specified} &
\multicolumn{2}{l}{\cellcolor[HTML]{EFEFEF} GPT-3 / Crowd} \\
\midrule
\textbf{Scenario} & \textbf{N} & \textbf{df} 
                 & \textbf{$\chi^2$}   & \textbf{p-value} 
                 & \textbf{$\chi^2$} & \textbf{p-value}
                 & \textbf{$\chi^2$} & \textbf{p-value}
                 & \textbf{$\chi^2$} & \textbf{p-value}
                 & \textbf{$\chi^2$} & \textbf{p-value}
                 \\ \midrule

Comm. compl.  & 1000 & 7      
& 30.25             & {\color[HTML]{9A0000} \textless{}.0001} 
& 37.03             & {\color[HTML]{9A0000} \textless{}.0001} 
& 10.65             & {\color[HTML]{9B9B9B}.15 (n.s.)}
& 10.57             & {\color[HTML]{9B9B9B}.16 (n.s.)}
& 32.16             & {\color[HTML]{9A0000} \textless{}.0001} \\

Content moder.       & 878 & 7      
& 28.61             & {\color[HTML]{9A0000} \textless{} .0001} 
& 35.70              & {\color[HTML]{9A0000} \textless{} .0001}     
& 7.39             & {\color[HTML]{9B9B9B}.39 (n.s.)}
& 4.95             & {\color[HTML]{9B9B9B}.67 (n.s.)}                  
& 49.42             & {\color[HTML]{9A0000} \textless{}.0001} \\

Disease diag.   & 920 & 7       
& 12.36             & {\color[HTML]{CE6301}.09} 
& 112.37             & {\color[HTML]{9A0000} \textless{} .0001}            
& 8.75            &  {\color[HTML]{9B9B9B}.27 (n.s.)}
& 10.47             & {\color[HTML]{9B9B9B}.16 (n.s.)}            
& 30.46             & {\color[HTML]{9A0000} \textless .0001}\\

Hiring   & 833 & 6      
& 14.33             & {\color[HTML]{CE6301} .03} 
& 58.69              & {\color[HTML]{9A0000} \textless{} .0001}            
& 5.05             & {\color[HTML]{9B9B9B}.54 (n.s.)}
& 5.83              & {\color[HTML]{9B9B9B}.44 (n.s.)}             
& 10.66              & {\color[HTML]{CE6301}.09 } \\

Loan applic.         & 751 & 5        
& 6.04              & {\color[HTML]{9B9B9B}.30 (n.s.)}                        
& 24.64             & {\color[HTML]{9A0000} \textless{} .0001} 
& 6.39             & {\color[HTML]{9B9B9B}.27 (n.s.)} 
& 2.67              & {\color[HTML]{9B9B9B}.75 (n.s.)} 
& 30.04             & {\color[HTML]{9A0000} \textless{}.0001} \\ \midrule
\end{tabular}
\vspace{-4pt}
\caption{$\chi^2$ statistics when comparing the distributions of harm categories across different experimental conditions.}
\label{table:chi-analysis-behaviors}
\vspace{-14pt}
\end{table}

\begin{figure}[t]
\includegraphics[trim= 0 0 0 20, clip,width=.9\linewidth]{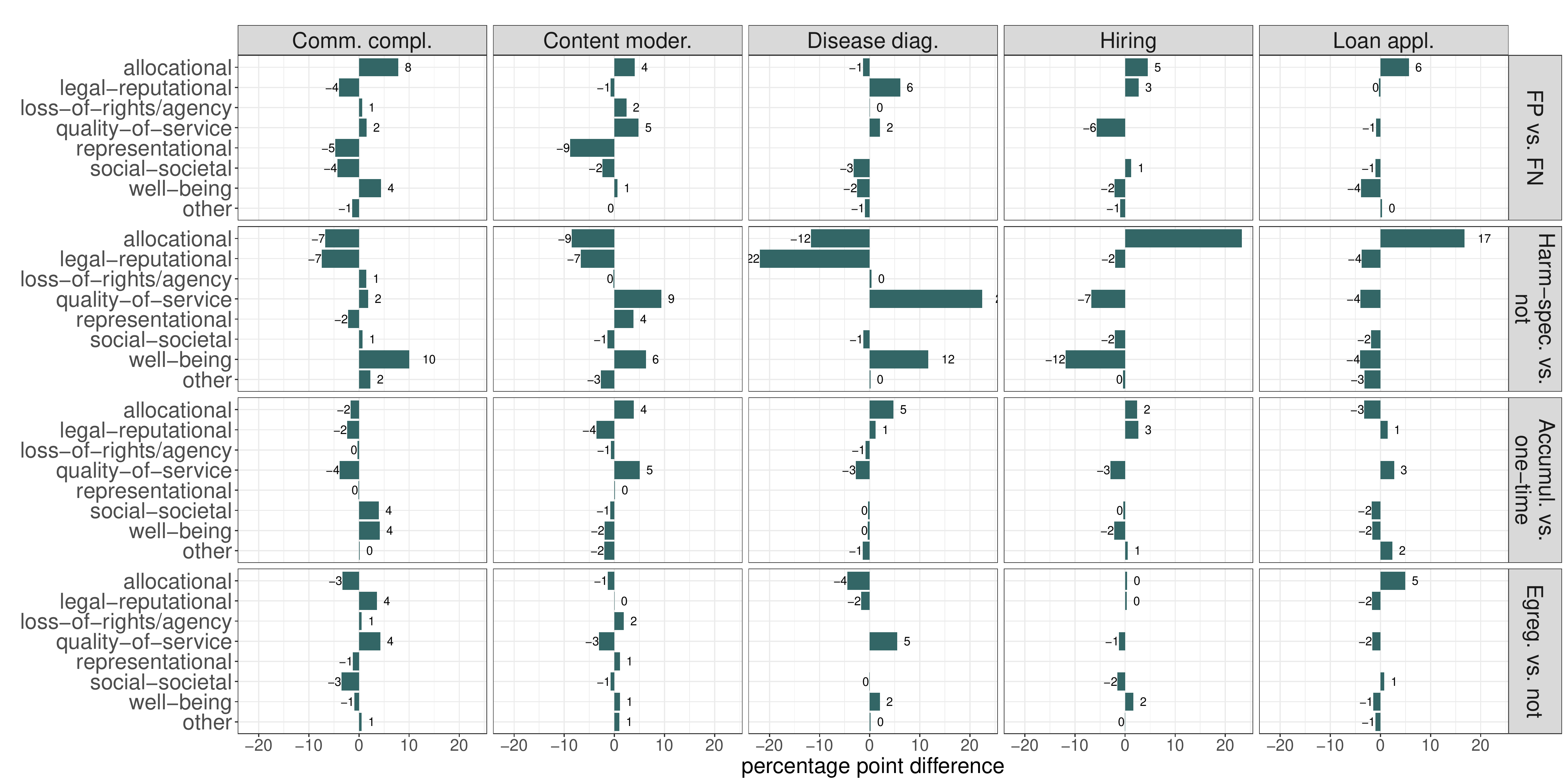}
\vspace{-12pt}
\caption{Percentage point differences in the fraction of different harm categories when contrasting AI behaviors across scenarios. When specifying the harm, we experimented with different harms across scenarios: \textit{emotional distress} for communication compliance and content moderation, \textit{financial concerns} for hiring and loan application, \textit{health concerns} for disease diagnosis.}
\label{figure:ai-behavior-per-scenario}
\vspace{-4pt}
\end{figure}

\begin{figure}[t]
    \centering
    \includegraphics[trim= 0 6 0 20, clip,width=\linewidth]{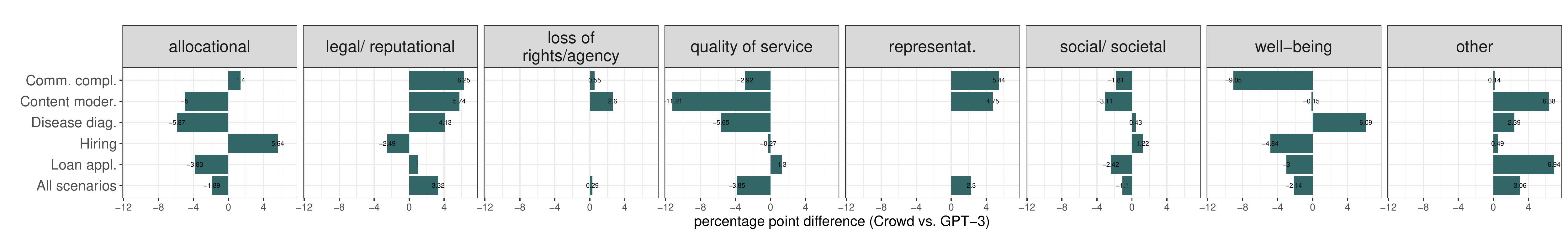}
    \vspace{-16pt}
    \caption{Distribution of harms surfaced by Crowd vs. GPT-3. Positive values indicate that Crowd, and negative values indicate that GPT-3 generated more examples of harms.}
    \label{fig:distribution-source-scenarios}
\vspace{-8pt}
\end{figure}

%\subsection{What does the systematic variation of problematic behaviors add to the generations?}
% H2: Does varying dimensions of problematic AI behaviors impact the harms generated?
\vspace{-2pt}
\subsection{RQ2: Does varying the dimensions of problematic AI behaviors impact the harms  \frameworkname{} surfaces?}
We manipulated various dimensions of problematic AI behaviors---{\em false positive/negative}, {\em accumulated/one-time}, {\em egregious/not-specified}, or {\em harm specified/not-specified}---in our experiments to understand their impact on harms generated. If priming on a behavioral dimension impacted harms surfaced by \frameworkname{}, we would expect to see differences in the types of harms generated across that dimension (e.g., between false positive vs false negative harms). To test this, we again ran Chi-square ($\chi^2$) analyses on the distribution of harm categories conditioned on each behavioral dimension.

Our results show that priming with {\em false positives vs. false negatives} behaviors resulted in significantly or marginally significantly different distributions of harm categories in each scenario, except for the loan application scenario (see Table~\ref{table:chi-analysis-behaviors}).
%{\em Priming with false positives vs. false negatives AI behaviors} resulted in significantly different distributions of harm categories, except for the loan application scenario (see Table~\ref{table:chi-analysis-behaviors}).
%Except for loan application, we see significant differences in the distributions of harm categories when priming with  behaviors (Table~\ref{table:chi-analysis-behaviors}). 
%From the figure, it can be seen that \textbf{different harms are accentuated for different types of AI mistakes} in different scenarios.
In the communication compliance and content moderation scenarios, for instance, false negatives (failure to detect toxic content) resulted in more examples of representational harms, while false positives (erroneously detecting that someone used toxic language) resulted in more examples of allocational  harms (see Figure~\ref{figure:ai-behavior-per-scenario}).
%Distinguishing between false positives and false negatives also resulted in different of harm categories for each stakeholder.

%\subsubsection{Does priming about specific harms help surface more examples of those harms?}
We also saw that priming about {\em specific harms} that practitioners may deem important for their scenario (e.g., you may experience \em emotional distress \em because ...) indeed helped surface more examples of those (or related) harms (see Figure~\ref{figure:ai-behavior-per-scenario}). For instance, we observed that priming about {\em financial concerns} resulted in a significantly higher prevalence of examples about allocational harms in both the hiring and loan application scenarios. 

We did not see any significant differences in distribution of harms when conditioning problematic behaviors on {\em one-time vs. accumulated} or {\em egregious vs. unspecified}. This could either be because our manipulations were not salient enough to elicit different harms for those dimensions or because those dimensions might not result in meaningfully different types of harm. For example, when varying the {\em one-time vs accumulated} dimension for users of a social media platform in the content moderation scenario, \frameworkname{} wrote "If the system determines a post contains toxic language when it does not, you may be harmed because.." in the {\em one-time} variation and "If the system determines posts contain toxic language when they do not, you may be harmed because.." in the {\em accumulated} variation. Future work might experiment with increasing the saliency of this dimension (e.g.,  "If the system often determines posts contain toxic language when they do not..").

\begin{figure}[t]
\includegraphics[trim= 0 15 0 0, clip, width = 0.7\linewidth]{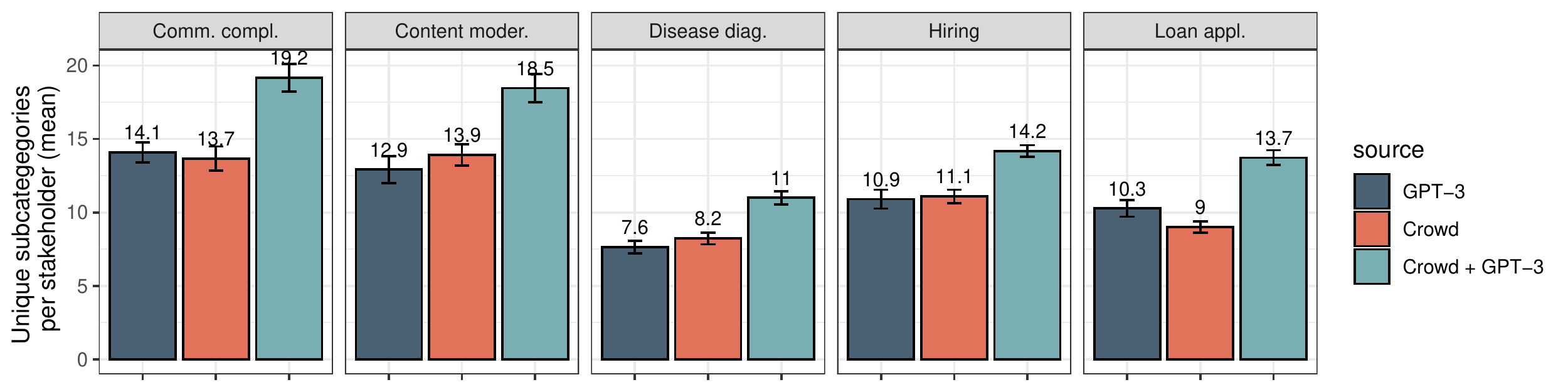}

\scriptsize
\begin{tabular}{p{0.7cm}|p{13.6cm}}
\hline
& \textbf{Crowd \& GPT-3}  
%& \textbf{Crowd only}  
%& \textbf{GPT-3 only}  
\\ \hline
\rowcolor[HTML]{EFEFEF} 
Comm.  compl. 
& backlash, bad actors, {\color{MidnightBlue}banned from site}, creating busywork, distrust and reputational damage, economic strain, {\color{MidnightBlue}eroding relationships}, feeling bad for others, job security, legal repercussions, liability, {\color{Bittersweet}loss of agency}, loss of motivation, loss of privacy, loss of rights, lost content, mental health, {\color{Bittersweet}opportunity loss}, {\color{MidnightBlue}physical health}, productivity loss, reinforced stereotype, safety, scapegoat, self-doubt, {\color{MidnightBlue}snowball effect}, target of toxic language, toxic environment, {\color{MidnightBlue}unaware}, underperforming AI, underspecified or repeated harm, unfair treatment, waste, {\color{MidnightBlue}work satisfaction/fit}  
%& %{\color{Bittersweet}loss of agency}, {\color{Bittersweet}opportunity loss}
%& %{\color{MidnightBlue}banned from site}, {\color{MidnightBlue}eroding relationships}, {\color{MidnightBlue}physical health}, {\color{MidnightBlue}snowball effect}, {\color{MidnightBlue}unaware}, {\color{MidnightBlue}work satisfaction/fit}
\\

Content moderat.   
& backlash, bad actors, banned from site, creating busywork, distrust  and reputational damage, economic strain, eroding relationships, feeling bad for others, {\color{MidnightBlue}inadequate service}, job security, lack of access to information, legal repercussions, liability, {\color{MidnightBlue}loss of agency}, {\color{Bittersweet}loss of motivation}, {\color{MidnightBlue}loss of privacy}, loss of rights, lost content, mental health, productivity loss, {\color{Bittersweet}public health}, {\color{MidnightBlue}opportunity loss}, reinforced stereotype, {\color{MidnightBlue}safety}, scapegoat, self-doubt, {\color{Bittersweet}social issues}, snowball effect, target of toxic language, toxic environment, unaware, underperforming AI, underspecified or repeated harm, unfair treatment, waste, {\color{MidnightBlue}work satisfaction/fit} 
%& %{\color{Bittersweet}loss of motivation}, {\color{Bittersweet}public health}, {\color{Bittersweet}social issues} 
%& %{\color{MidnightBlue}inadequate service}, {\color{MidnightBlue}loss of agency}, {\color{MidnightBlue}loss of privacy}, {\color{MidnightBlue}opportunity loss}, {\color{MidnightBlue}safety}, {\color{MidnightBlue}work satisfaction/fit} 
\\

\rowcolor[HTML]{EFEFEF} 
Disease\quad diag.    
& .{\color{MidnightBlue}bad actors}, creating busywork, {\color{Bittersweet}commiseration}, distrust  and reputational damage, economic strain, {\color{MidnightBlue}eroding relationships}, inadequate service, job security, legal repercussions, liability, {\color{MidnightBlue}loss of agency}, {\color{MidnightBlue}loss of privacy}, mental health, physical health, public health, {\color{MidnightBlue}quality of life}, {\color{Bittersweet}reinforced stereotype}, scapegoat, snowball effect, underperforming AI, underspecified or repeated harm, unfair treatment, waste               
%& %{\color{Bittersweet}commiseration}, {\color{Bittersweet}reinforced stereotype}                                                               
%& {\color{MidnightBlue}bad actors}, {\color{MidnightBlue}eroding relationships}, {\color{MidnightBlue}loss of agency}, {\color{MidnightBlue}loss of privacy}, {\color{MidnightBlue}quality of life}                        
\\

Hiring     
& .{\color{Bittersweet}bad actors}, creating busywork, {\color{Bittersweet}commiseration}, distrust  and reputational damage, economic strain, eroding relationships, job security, legal repercussions, liability, loss of motivation, mental health, opportunity loss, productivity loss, {\color{Bittersweet}reinforced stereotype}, scapegoat, self-doubt, snowball effect, toxic environment, underperforming AI, underspecified or repeated harm, unfair treatment, waste, work satisfaction/fit 
%& %{\color{Bittersweet}bad actors}, {\color{Bittersweet}commiseration}, {\color{Bittersweet}reinforced stereotype}  
%& %-      
\\

\rowcolor[HTML]{EFEFEF} 
Loan \quad \quad appl.          
& .{\color{MidnightBlue}bad actors}, creating busywork, {\color{Bittersweet}commiseration}, distrust  and reputational damage, economic strain, eroding relationships, incapability, job security, legal repercussions, liability, mental health, opportunity loss, productivity loss, quality of life, scapegoat, {\color{MidnightBlue}self-doubt}, {\color{Bittersweet}snowball effect}, social issues, underperforming AI, underspecified or repeated harm, unfair treatment, {\color{MidnightBlue}waste}    
%& %{\color{Bittersweet}commiseration}, {\color{Bittersweet}snowball effect}                                                                
%& %{\color{MidnightBlue}bad actors}, {\color{MidnightBlue}self-doubt}, {\color{MidnightBlue}waste}                                                                                                            
\\ \hline
\end{tabular}
\vspace{-10pt}
\caption{Unique harm subcategories covered by the examples obtained for each \em scenario \em with Crowd and GPT-3.
The top plot shows the means of unique number of subcategories \em per stakeholder \em generated by {\color{Bittersweet}Crowd} only, {\color{MidnightBlue}GPT-3} only, and Combined (Crowd + GPT-3). Error bars indicate one standard error. %Crowd and GPT-3 combined generate significantly more unique subcategories per stakeholder than either Crowd or GPT-3 alone. 
(See Table~\ref{table:complementarity} for significance tests.)
}
\label{table:diff-unq-codes}
\vspace{-8pt}
\end{figure}

\vspace{-4pt}
%qualitative examples of diverse harms%
\subsection{RQ3: Does eliciting harms from crowds vs. GPT-3 impact the harms generated?}
\label{subsec:crowd-gpt3}
\frameworkname{} generates vignettes that set up fictional scenarios to be completed with possible harms. In our experiments, we tried prompting crowdworkers and GPT-3 to complete these vignettes to understand the trade-offs between these approaches. 

To examine the trade-offs, we first compared the distribution of harm categories surfaced by crowds versus GPT-3 as before with Chi-square ($\chi^2$) analyses (see Table~\ref{table:chi-analysis-behaviors}). We found significant differences in the distribution of harms across harm categories between crowds vs. GPT-3 in all scenarios, with the exception of the {\em hiring} scenario where we found marginally significant differences. 
When examining these differences (see Figure ~\ref{fig:distribution-source-scenarios}), we see that overall GPT-3 generated more examples of e.g., {\em quality of service}, {\em allocational}, {\em well-being} harms, while the crowds provided more examples of e.g., {\em legal \& reputational} and {\em representational} harms. 

%We examined differences in the prevalence and diversity of the harm examples elicited from either crowds or GPT-3 to understand how they complement each other, if at all. The distribution of examples across harm categories was significantly different between crowds and GPT-3 (see Table~\ref{table:chi-analysis-behaviors}). 

Next, we wanted to compare crowds vs. GPT-3 in terms of the diversity of harms surfaced by each approach. As a proxy for diversity, we look at the total number of unique harm subcategories each approach surfaced across their stakeholder groups. That is, because the same harm subcategory can have widely different consequences depending on which stakeholder they affect (e.g., %in disease diagnosis, 
reputational harms towards the whole medical community vs. a single doctor have vastly different impacts) %~\footnote{Illustrating this point, harms generated by AHA! for the medical community as a stakeholder include: ``the reputation and validity of all the other practitioners may be questioned'' vs. the doctor: ``I will lose my reputation.''} 
we count these separately.\footnote{Note that this may still be considered a lower bound measure of diversity because two harms with the same subcategory for the same stakeholder could also represent different consequences (e.g., ``Then despite my best efforts, I might then be relieved soon of duty, and would then have a difficult practical time of it to gain/regain unemployment benefits from my state, when it might have been better that no hiring of me happened in the first place due to the faulty AI'' and ``your salary might be lower than it should be'' ).} %Also since a single subcategory can and does encapsulate multiple distinct harms, the number of unique subcategories captures only partially the diversity of the generated harms.} 
We conducted paired t-tests comparing the mean number of unique subcategories of harms generated by Crowds vs GPT-3 and between Crowds-only, GPT-3-only, and their Combination (Crowd + GPT-3) (see Table~\ref{table:complementarity}). We found that while using either only crowds or only GPT-3 resulted in a comparable number of unique harm subcategories, together they produce a more comprehensive set of subcategories in all scenarios which are both significantly larger and significantly more diverse than either GPT-3 or crowd alone (see Figure~\ref{table:diff-unq-codes}).

\begin{table}[h]
\scriptsize
\centering
\begin{tabular}{@{}lp{1.4cm}lllll}
\toprule
\rowcolor[HTML]{EFEFEF} 
Scenario    
& \begin{tabular}[c]{@{}l@{}}Crowd mean \\ (standard error)\end{tabular} 
& \begin{tabular}[c]{@{}l@{}}GPT-3 mean \\ (standard error)\end{tabular} 
& \begin{tabular}[c]{@{}l@{}}Crowd + GPT-3 \\ (standard  error)\end{tabular} 
& \begin{tabular}[c]{@{}l@{}}Crowd vs. GPT-3\end{tabular}                                     
& \begin{tabular}[c]{@{}l@{}}Crowd vs. \\ Combined\end{tabular}                               
& \begin{tabular}[c]{@{}l@{}}GPT-3 vs. \\ Combined\end{tabular}                               
\\ \midrule

\textbf{\begin{tabular}[c]{@{}l@{}}Comm. compl.\end{tabular}} & \begin{tabular}[c]{@{}l@{}}13.67 (0.83)\end{tabular}                     & \begin{tabular}[c]{@{}l@{}}14.08 (0.68)\end{tabular}                     & 19.17 (0.93)                                                                                             & {\color[HTML]{9B9B9B} \begin{tabular}[c]{@{}l@{}}t(11) = -0.5 p = .63 (n.s.)\end{tabular}} & {\color[HTML]{9A0000} \begin{tabular}[c]{@{}l@{}}t(11) = -11 p < .0001\end{tabular}}    & {\color[HTML]{9A0000} \begin{tabular}[c]{@{}l@{}}t(11) = -8.53 p < .0001\end{tabular}}    \\

\textbf{Content moder.}                                                 & \begin{tabular}[c]{@{}l@{}}13.92 (0.72)\end{tabular}                     & \begin{tabular}[c]{@{}l@{}}12.92 (0.92)\end{tabular}                     & 18.46 (0.97)                                                                                            & {\color[HTML]{9B9B9B} \begin{tabular}[c]{@{}l@{}}t(12) = 1.85 p = .13 (n.s.)\end{tabular}}  & {\color[HTML]{9A0000} \begin{tabular}[c]{@{}l@{}}t(12) = -8.42 p < .0001\end{tabular}}     & {\color[HTML]{9A0000} \begin{tabular}[c]{@{}l@{}}t(12) = -10.29 p < .0001\end{tabular}}    \\

\textbf{Disease diag.}                                                  & \begin{tabular}[c]{@{}l@{}}8.23 (0.40)\end{tabular}                      & \begin{tabular}[c]{@{}l@{}}7.65 (0.42)\end{tabular}                      & 11.00 (0.45)                                                                                                & {\color[HTML]{9B9B9B} \begin{tabular}[c]{@{}l@{}}t(16) = 1.61 p = .13 (n.s.)\end{tabular}}  & {\color[HTML]{9A0000} \begin{tabular}[c]{@{}l@{}}t(16) = -8.76 p < .0001\end{tabular}} & {\color[HTML]{9A0000} \begin{tabular}[c]{@{}l@{}}t(16) = -10.87 p < .0001\end{tabular}} \\

\textbf{Hiring}                                                             & \begin{tabular}[c]{@{}l@{}}11.09 (0.46)\end{tabular}                     & \begin{tabular}[c]{@{}l@{}}10.91 (0.64)\end{tabular}                     & 13.58 (0.40)                                                                                            & {\color[HTML]{9B9B9B} \begin{tabular}[c]{@{}l@{}}t(10) = 0.27 p = .79 (n.s.)\end{tabular}}  & {\color[HTML]{9A0000} \begin{tabular}[c]{@{}l@{}}t(10) = -6.49 p < .0001\end{tabular}}     & {\color[HTML]{9A0000} \begin{tabular}[c]{@{}l@{}}t(10) = -9.11 p < .0001\end{tabular}}     \\

\textbf{Loan applic.}                                                   & \begin{tabular}[c]{@{}l@{}}9 (0.38)\end{tabular}                         & \begin{tabular}[c]{@{}l@{}}10.27 (0.56)\end{tabular}                     & 13.72 (0.51)                                                                                            & {\color[HTML]{9B9B9B} \begin{tabular}[c]{@{}l@{}}t(10) = -2.05 p = .07 (n.s.)\end{tabular}} & {\color[HTML]{9A0000} \begin{tabular}[c]{@{}l@{}}t(10) = -11.63 p < .0001\end{tabular}} & {\color[HTML]{9A0000} \begin{tabular}[c]{@{}l@{}}t(10) = -8.37 p < .0001\end{tabular}}    \\ \bottomrule
\end{tabular}
\caption{Paired t-test comparing the mean number of unique subcategories of harms generated by Crowd only, GPT-3 only, and Combined (Crowd + GPT-3). For all the scenarios, Crowd and GPT-3 combined generate significantly more unique subcategories together than either alone.}
\vspace{-8pt}
\label{table:complementarity}
\end{table}

\section{Practitioner Perspectives on \frameworkname{}}
\label{sec:qual}
To assess the potential practical utility of \frameworkname{}, we conducted semi-structured interviews with industry practitioners and academics with responsible AI expertise. This section describes our IRB-approved study and findings. 
%Prior to conducting it, the interview study was IRB-approved. In the remaining of the paper, anonymized participant quotes---edited and paraphrased for brevity and clarity---are highlighted in italics followed by a "P" and an id.

\subsection{Interview Study}

\noindent\textbf{Participants. }
%We recruited responsible AI (RAI) experts to evaluate our framework. 
We reached out to $20$ practitioners and academics with responsible AI experience (e.g., experience conducting or reviewing AI product impact assessments) or expertise from our professional networks and through snowball sampling. Nine people accepted our invitation, including seven from a large technology company and two academics from different institutions. Participants covered four professional roles: {\em Project Managers} [P1, P7]---project managers with specific training in responsible AI concepts and best practices and with experience helping development teams conduct impact assessments, {\em Responsible AI Practitioners} [P4, P5, P6]---practitioners whose work involves researching, guiding, or developing responsible AI tools and practices, {\em Responsible AI Policy} [P8, P9]---practitioners responsible for development of and compliance with responsible AI policies, and {\em Responsible AI Academics} [P2, P3]---university professors who study responsible AI challenges and solutions.

\para{Protocol.}
Before each interview, we obtained informed consent from participants and asked for their permission to record the session. Interviews were conducted via a video call platform. We started each interview with an overview of the project and study protocol. Each interview then consisted of two tasks featuring two distinct AI deployment scenarios randomly chosen for each participant from our five original scenarios. For each task, participants were first presented with an AI deployment scenario and then asked to freely brainstorm possible harms. They were then shown the harms surfaced by \frameworkname{} for that scenario via a simple interactive tool we designed to help them navigate the results (see Appendix~\ref{appendi:interface}). The tool grouped harms by stakeholder and harm categories and subcategories. Participants could explore harms by filtering them by stakeholder or expanding harm categories to reveal individual harms. %
We asked participants to think aloud throughout the interview and reflect on anything notable about specific harms \frameworkname{} surfaced. We also asked for participants' reflections on \frameworkname{}'s overall approach and their thoughts on its potential practical utility. Each interview lasted about 50 minutes with task 1 taking most of the time (typically 30-35 minutes) while task 2 was used solely as an overview of a different scenario for the remainder of the time (about 5-7 minutes).

\para{Qualitative analysis.}
To analyze the interview data, we used a bottom-up thematic analysis approach ~\cite{strauss1994grounded}. First, we reviewed transcripts and videos highlighting quotes relevant to our study goals of understanding participant reactions to \frameworkname{}'s generated harms, overall approach, and practical utility. We then thematically sorted excerpts to identify themes. 

% Discussion and future work
% Coverage
    % Combining humans and LLMs lead to more diversity (and different harms?)
    % P2 mentioned something about this?

% Overreliance, changing the problem. Could be used by reviewers, for a database for learning, etc. Presentation improvements.
    % how to present the harms -> problem changes from no harms to too many harms. 
    % Preventing overreliance
    % Other uses -> as a repository for people to learn from
    % how to present the harms -> careful on counts, misleading, future work

% Supporting decision making -> its the first step, need to connect with severity assessment and then mitigations
    % Supporting informed decision making -> severity, mitigation, thresholds -> future work
    % P3 mentioned use for giving evidence to decision makers

% Harms missing. Probelmatic behaviors, future work.
    % focused on only problematic behaviors -> future work
    % Platform systems -> need to contextualize. Empathy. % P4, most of the times, teams will go for hte most obvious
    % Eliciting empathy helped, from pilot studies
    % More details about the context helped, from pilot studies. Difficult for platform systems. Future work.
    % P9 use for researchers and education

%Limitations -> overreliance concerns, biases of language models

\subsection{R4: Do experienced professionals see value in \frameworkname{} for helping teams anticipate harms in practice?}
\para{\frameworkname{} surfaced sensible harms participants believed they otherwise would not have thought of.} All our participants pointed out harms surfaced by \frameworkname{} that they found sensible but unlikely to have thought of on their own. When reviewing harms towards direct stakeholders (e.g., decision subjects) in the communication compliance (in the workplace) scenario, P8 said \textit{``[What] was surprising to me was loss of motivation. [There was] an article from the New York Times [about workplace monitoring]. [T]his is just a form of workplace monitoring, right? [T]he article is more about like productivity [and] tracking how much you're like online. [This is] kind of a different version of that too.''} 
In another example, P2 commented about mental health related harms towards applicants in the loan application scenario, saying \textit{``Mental health, that's surprising to me because that's one I think of as a risk for other applications, particularly in the online setting and I guess here [in] a loan application scenario [..] it's more of a second order effect from having a loan denied [..] that's one I wouldn't have immediately thought of in this context.''}
Similarly, when reviewing possible harms towards indirect stakeholders in the content moderation scenario, P6 remarked \textit{``I would not have thought about content moderators. And I should have thought about them because I know that there's a lot of stress, it's a very hard job.''} In another example from this scenario, P3 said \textit{``[I]t's interesting how people think that AI system developers could be fired for the performance of the systems''} when reviewing possible harms surfaced towards system developers. 

Although all participants pointed out surprising harms they considered sensible, some also noted harms they believed to be out of scope. % or should be considered separately by development teams. 
When examining emotional harms that applicants could face in the hiring scenario, P3 said \textit{``emotional kind of harm, that I wouldn't necessarily associate with the system, but just the nature of hiring processes.''} 
Another participant argued that harms caused by misusing AI systems are different than downstream harms from intended uses. Specifically, in the context of discussing harms of misuse surfaced by \frameworkname{} in the communication compliance scenario (e.g., \textit{``[s]omebody has used someone else's laptop or their e-mail account to send [toxic] emails in order to try to get them fired''}~[P4]), P4 said \textit{``[T]here's appropriate use of the system %So whoever is the end user interacting with it, they're using it as it was designed. 
[..] There is unsupported use where the system wasn't designed to work this way [..] And then there is misuse, [..] somebody is actively going into this with like adversarial malicious intent [..] I think that is distinct from sort of the downstream harms that might occur.''} 
%In Section X, we discuss the tradeoffs between a comprehensive approach to harm anticipation versus other considerations such as review costs and mitigation burden.

    % Any comments on how these might be hard for teams to think of?
    % Any idea if the things they mentioned in the initial brainstorms were covered by Aha?

%Indeed some of the descriptions of harms generated by our framework may be not directly related to the system and arise when a wrong decision is made by any classifier, including a human. We argue, however, that having an AI system make the decisions instead of a human, for example, may often exacerbate some of these harms. For instance, a wrongfully denied loan application may be more easily appealed if resulting from a human decision-maker compared to a human aided by an (often uninterpretable) AI, for which recourse pathways are more difficult if at all present~\cite{}.

\para{Participants noted that \frameworkname{}'s systematic approach was important for considering a broad range of harms.} 
Several participants commented that \frameworkname{}'s systematic approach was important for increasing the coverage of harms considered when developing AI systems. P7, a PM with experience helping product teams consider the potential impact of their AI systems, said \textit{``One of the challenges we have is that it is not clear how well we have covered all of the potential harms. [S]o it is regularly the case that we'll go through impact review, maybe some members of the V team aren't there because they had conflicts. Maybe people are just having an off day and they're not being very creative or thoughtful [..] there are all these human opportunities for error, so having a tool in place that helps us to be more systematic is really essential.''}

Three participants highlighted specific components of \frameworkname{} as impactful. P2 called out the value of thinking broadly about stakeholders when examining harms: \textit{``Other applicants is one that's very easily forgotten about [in the hiring scenario]. And so that's nice that that's there as a reminder. And family and friends of the applicant, that's probably one that's overlooked.''}  
P5 stressed the benefits of drawing attention to different problematic behaviors: \textit{``in [the communication compliance] scenario, I think it's easy to think about harms that can arise with false positives, because there are all these reputational damages and potential damages to job security. But thinking about false negative harms may be a little bit more challenging.''} Others noted the value of connecting stakeholders to system behaviors, with P1 saying \textit{``[it] has a really good structure and having just read a very structured approach to generating harms made it much easier to repeat that [like] starting with stakeholders, listing types of harms and then connecting types of harms to stakeholders.''} 

Interestingly, before participants first saw \frameworkname{}'s results, most appeared to take some kind of structured approach when asked to brainstorm harms on their own. P5, P6, and P7 began listing harms that arise due to problematic AI behaviors such as false positives and false negatives, while P8 and P9 spoke to problematic AI behaviors in connection with stakeholders. While this might be an artifact of their prior training or experience, it also provides some face validity to the benefits of \frameworkname{}'s approach given that all of our participants had some level of Responsible AI expertise. 
%Moreover, all participants noted valid harms surfaced by \frameworkname{} that they believed they would not have otherwise anticipated, suggesting \frameworkname{}'s explicit and broad coverage of stakeholders and behaviors provided added value.
%
Of the remaining participants, P1, P2, and P4, approached harm anticipation by considering violations to common responsible AI desiderata (e.g., privacy, fairness) while P3 started with a high-level taxonomy of harms they tried to map to the deployment scenario they were presented with. This could be considered a top-down approach to anticipating harms, whereas \frameworkname{} generates descriptions of harms which can then be grouped into harm categories bottom-up. However, the \frameworkname{}'s ethical matrix could be extended to condition on responsible AI desiderata or harm taxonomies, similar to how we incorporated harm specifications in our experiments.

\para{Participants differed in their opinions about when and how \frameworkname{} should or could be used in practice.} %Largely determined by their role, participants expressed different ideas as to when in the ideation process would our framework be helpful and which decision-makers may find it most useful.
%Participants expressed different opinions about when and how a system like \frameworkname{} should or could be used. 
Participants who had experienced teams struggling with conducting impact assessments commented on the benefits of using a tool like \frameworkname{} as a starting point. P5 noted that \textit{``[product] teams are struggling with starting from basically zero and do not have any training in responsible AI to anticipate these types of harms,''} adding that a tool like this \textit{``would give them a place to start.''} Similarly, P8 commented that \textit{``this would be better for like the teams who are actually filling out the impact assessment [like] kind of [a] starting place for brainstorming.''} 
%P9 also comments about not ahving to start from scratch, particularly for the platform scenario

Other participants recommended that a tool like \frameworkname{} should be used during brainstorming to broaden perspectives and coverage of harms. P1 commented that  \textit{``I think it's useful both as like a cross check. Is there anything that I missed [sic] general brainstorm that can then be taken to a deeper consideration of different kinds of harms. Also definitely useful to get additional perspectives here, thoughts about harms, especially for people who are maybe approaching this kind of thought process for the first time.''} P6 similarly remarked that teams could use \frameworkname{} \textit{``when they have that brainstorming block [and] can't think of any other things [..] helping with broadening that brainstorming field.''} 
% overreliance could go here

%P1: ...the system is doing a great job at breath, so I personally would be interested in using a system like this, maybe in the context of a review of an AI feature to make sure that we haven't missed anything in conversation. Let's brainstorm a few harms, or stakeholders and then take a look at what are some other options and maybe there will be some good pieces there and maybe there will be.

Still others advocated for using tools like \frameworkname{} only after teams deeply engage with the task themselves. P4 said \textit{``I would want to continue using this as sort of that secondary check because I do like forcing teams into some open world envisioning and before they have a framework that they can just lean on and check boxes and look at specific things and call it a day.''}  Similarly, P8 pondered about whether \textit{``it is better to have a completed impact assessment where somebody has copy-pasted [from the generations] or have practitioners think through these things and learn to flex this muscle.''} % or here
%In Section \ref{sec:discussion}, we reflect on these differing opinions and on how to leverage the benefits of automatically surfacing meaningful harms while encouraging practitioners to deeply engage in this important and consequential task. 

%Product PMs (P1, P7)---who have conducted impact assessments along with their teams---advocated using a tool like this as a starting point the tool being used early in the ideation process, with P7 noting ``the idea here is to help teams [and] whoever is doing this task of thinking of harms to not start with a blank page [to] help people come up with things that they haven't thought of before.'' 

\para{Participants suggested various ways that \frameworkname{} could or should inform development decisions.} 
Particpants recognized ways in which \frameworkname{} could already inform development decision-making. P5 noted that looking at the harms caused by different problematic AI behaviors might directly imply some mitigations:  \textit{``maybe this tool can help people think about [..] what kind of mistakes lead to more harms? Is it false positives or is it false negatives? So that [it] can help developers [..] focus their attention on making specific improvements to the model.''} 
Other participants saw the comprehensive set of harms produced by \frameworkname{} as providing convincing evidence that could encourage leaders and decision-makers to exercise caution or better resource mitigation work. P4 noted that \textit{``[W]here I see this being useful is if I need to convince someone. Basically, you need to be really worried about this because look at all of the things that might go wrong that we feel really uncomfortable with having a long laundry list of things can somewhat help with landing that message.''} Similarly, P3 said \textit{``OK well I'm creating this system. How big of a trouble am I in? Let me quantify it [and then] talk to an executive and say Hi, look at this like we've got 700 issues on the legal consequences of this [..] maybe the executives will say yes, I will give you extra budget for the legal team for this project.''}  

% Others wanted more support for actionability
Other participants wanted more explicit support for mitigations. % in \frameworkname{}~ 
P9 remarked \textit{``it's a really hard thing to assess severity [so] some indication of the severity of harm [..] would be really useful [I'd like to] click on a harm and have a suggested mitigation.''} They further elaborated that \textit{``we're asking our product teams to make choices about which harms they mitigate based on limited resources and [..] need more help on like is this above the line or below the line? It's great to know this stuff will happen, but what are we willing to sign up to mitigate?''} Similarly, P5 said \textit{``[W]hen people look at harms, they also want to think about whether there is an opportunity for mitigation.''}

\vspace{-2pt}
\section{Discussion \& Future Work}
\label{sec:discussion}

%shift harm anticipation from an ideation problem to a potentially demanding review problem

\para{Trade-offs between a harm ideation problem versus a potentially demanding review problem.} 
\frameworkname{} was designed to support practitioners tasked with anticipating a wide range of downstream harms, a task responsible AI proponents advocate for to mitigate the negative impact of AI-based systems to people and society~\cite{nanayakkara2021unpacking,boyarskaya2020overcoming,smith2022real}. Our experiments and interview study show that \frameworkname{} can help surface a wide range of meaningful harms that even experienced responsible AI practitioners believed they would have likely overlooked. However, several practitioners were also concerned that \frameworkname{} may be shifting the problem from thinking of harms to reviewing a large number of potentially noisy examples. P8 remarked \textit{``If I put on my value sensitive design kind of hat, yeah, absolutely [..] let's think about and legitimate all these stakeholders. When I think about like doing this as a practitioner [..] the flip side is with a tool like this you might generate so many different harms that the work is actually going through and filtering out what are the actionable harms?''} 
While we believe \frameworkname{} can be extended to help prioritize the harms it helps surface (e.g., by recruiting crowd judges to rate the potential severity of the harm examples as a subsequent stage), we recommend caution when automating value-laden decisions such as what harms to prioritize mitigating. 

% P2 Multiple participants noted that a long list of generations can be \textit{``overwhelming to a developer who just sort of first wants to understand what the things are that they should care about''} and they would prefer to be presented with \textit{``a sort of a summary.''} 
% P7: %The the onus is on both the feature crew and the RAI V team that we have to to sort of look at the scenario and generate those harms and the risk that you run there is that. You know, amongst that crowd of individuals, you may you may not identify all of the harms right. It's the there's obviously human error there, but you know, the more individuals we have involved, the better you're gonna be at trying to cover all the bases. 

%\para{Trade-offs between automatically surfacing harms versus encouraging practitioners to deeply engage in the task.} 
\para{Trade-offs between offloading harm anticipation to a tool like \frameworkname{} versus encouraging practitioners to deeply engage in the task.} 
Several of our study participants expressed concerns that a tool like \frameworkname{} could in fact encourage practitioners to sidestep the task of examining the impact of their technologies or could disincentivize them from learning to consider the consequences of their technologies. 
We acknowledge this trade-off, and recommend \frameworkname{} be primarily used as a secondary check to first allow on-the-ground practitioners the opportunity to \textit{``learn to flex this muscle''}~[P8], as some of our participants suggested. 
%or used by reviewers or auditors (e.g., [X,X]) to leverage \frameworkname{}'s benefits of harm anticipation while  
Future work may also consider using \frameworkname{} as an educational tool or to grow a repository of harms that arise across various AI deployment scenarios for knowledge sharing and further study. P9 emphasized \textit{``nobody should be starting from scratch with an impact assessment [..] we should be building on each other's learning and understanding what's different about the nuances of my scenario or my system, not the things that we already know. You know, what's the delta from what we do know?''} 

% overreliance, CYA
%Participants thought the risk of overreliance also makes it critical \textit{``to communicate that [the tool] is not perfect, [that] there are things that these generations may overlook, so teams still should do the thinking around these harms themselves and use [the tool] as a starting point''}~[P5]. 

\para{Tradeoffs between decoupling versus tying harm anticipation to mitigations.} Participants, particularly those tasked with implementing responsible AI mitigations, wanted more guidance on actionability, with P4 asking \textit{``if I'm approaching this from a product perspective and I'm figuring out how do I actually build this system responsibly, what are the things I need to mitigate? What do I do with this information?''} %and P5 commenting \textit{``that teams would need to discuss the types of harms that can arise from the AI system they're trying to build with the leadership [including] which of those harms they think are addressable especially in the timeline that they have.''}
While we believe \frameworkname{} can be extended to provide more guidance on actionability, we argue that decoupling harm anticipation from harm mitigation is critical for several reasons. First, research on effective brainstorming strategies emphasize the benefits of divergence before convergence~\cite{finke1996creative}. We took this approach with \frameworkname{} to encourage open and unrestricted consideration of a wide range of harms before fixating on any particular problem or solution. Some participants indicated they appreciated this separation. P8---a responsible AI policy practitioner with extensive experience reviewing and guiding teams on impact assessments---noted \textit{``[teams] tend to be really focused on problem-solving. So a lot of times, the harm comes with the solution [..] Our process is like trying to decouple those things, right? Like first you think about all the things that can go wrong and then eventually you move toward, how do you fix this?''} 
Second, while automating part of the generation of possible harms can increase coverage of harm considerations, we believe the community should exercise caution when attempting to automate value-laden decision-making such as harm prioritization. P6, for example, pushed back on the idea of assigning severity scores to harms asking \textit{``who assigns the severity scores''} and \textit{``whose values and whose perspective are reflected [by the severity scores].''} 
Finally, knowledge for how to mitigate the harms surfaced by \frameworkname{} may require deep organizational, technical, and regulatory expertise, which remains difficult to reliably provide especially in an semi-automated fashion or at scale.

\para{Tradeoffs when eliciting harms from crowd judges and GPT-3.} In its' current implementation, \frameworkname{} elicits examples of harm from both crowd judges and GPT-3 to uncover a wider range of issues by providing perspectives beyond those held by often homogeneous AI teams~\cite{boyarskaya2020overcoming,ONeil2020NearTermAI,Ballard2019JudgmentCT} and make it easier to conduct impact assessments. 
While in our experiments, using both crowd judges and GPT-3 overall resulted in more diverse and comprehensive examples of harm (Section~\ref{subsec:crowd-gpt3}), and while some participants recognized that trying to provide more diverse perspectives is \textit{``the right way of thinking about how to develop [tools like \frameworkname{}]''}~[P2], the examples from both the crowd judges and GPT-3 may still reflect the same `status quo' and may surface harmful biases~\cite{wagner2021measuring}. For instance, crowd judges heavily skew towards certain demographics (young and white)~\cite{goodman2017crowdsourcing}, while LLMs were found to produce language that stereotypes, demeans, and erases groups, individuals, or their experiences~\cite{hartvigsen-etal-2022-toxigen,gehman-etal-2020-realtoxicityprompts,sheng2019woman,weidinger2022taxonomy,bender2021dangers}. %These issues may thus end up being reflected in how our vignettes are being completed. 
To mitigate some of these, future work could draw on creativity research~\cite{siangliulue2016ideahound, rhys2021directed} and, for instance, explore different approaches to eliciting descriptions of possible harms by priming crowd judges (and LLMs) with different types of harm examples to boost or guide their imagination.
Furthermore, \frameworkname{} can also be extended to include other ways to complete the vignettes e.g., by having the responsible AI practitioners complete the vignettes by themselves, or by recruiting diverse participants representing relevant stakeholders in a given scenario.

\vspace{-4pt}
\subsection{Limitations \& Ethical Considerations}
While our goal is to facilitate ethical reflection, as we discussed in the previous sections, tools like \frameworkname{} also come with risks, especially due to how they might be used (e.g., people might over rely on \frameworkname{} to superficially comply with responsible AI requirements), or due to how the examples of harm were obtained (e.g., they might surface harmful biases encoded by LLMs or held by crowd judges). 
Furthermore, in its' current instantiation, \frameworkname{} is designed to surface harms dues to problematic AI behaviors. However, harms may arise not only due to unintended, problematic AI behaviours; but also due to intended uses particularly when AI systems are employed for tasks that are conceptually or practically impossible (e.g., predicting criminality from physical appearance)~\cite{raji2022fallacy}, or due to malicious uses such as employing AI systems for undemocratic or fraudulent purposes~\cite{weidinger2022taxonomy}. 
By attempting to formalize and automate parts of the process of ideating about possible harms (particularly with the use of LLMs), our approach might also be critiqued as ``algorithmic formalism''~\cite{green2020algorithmic}. 
We, however, envision \frameworkname{} being used more as a secondary check or to support practitioners when they do not know where to start, and we do not recommend using it as standalone to fill out impact assessment---which we believe accounts for its' limitations and the realities practitioners need to navigate. The design of \frameworkname{} also recognizes that the process of envisioning harms is inherently contextual.

\balance{}

\iffalse
\section{Acknowledgements}

FATE, Hal Daume III, Su Lin Blodgett, Mahsan Nourani, Krzysztof Gajos, Stephanie Ballard, pilot interviewees
\fi

\bibliographystyle{SIGCHI-Reference-Format}
\bibliography{references}
\newpage
\appendix
\section{Appendix}

\subsection{Scenario Descriptions}
In this Appendix, we provide additional details about the scenario descriptions we used in our experiments (Table~\ref{ref:scenario-description}), including the stakeholders (Table~\ref{table:stakeholders}). % and example vignettes (Table~\ref{tab:sample_vig}). 

\begin{table}[h]
\scriptsize
\begin{tabular}{p{2.5cm}|p{12.5cm}}
        & \textbf{Scenario description} \\ \hline

\rowcolor[HTML]{EFEFEF} 
\textbf{Communication  compliance} 
& A company wants to deploy an AI system to scan its employees' emails for toxic language. If the system detects that an employee is using toxic language in their emails,  it will automatically notify the employee's manager. \\

\textbf{Content  moderation}                               
& A social media company wants to deploy an AI system to detect and remove posts that may contain toxic language.    \\

\rowcolor[HTML]{EFEFEF} 
\textbf{Disease diagnosis} 
& A medical clinic wants to deploy an AI system to scan the medical records of patients and determine whether they have a certain disease. \\

\textbf{Hiring}
& A tech company wants to deploy an AI hiring system to scan the resumes of applicants and predict whether they are a good fit for a given job opening.  \\

\rowcolor[HTML]{EFEFEF} 
\textbf{Loan application}          
& A bank wants to deploy an AI system to scan the information of loan applicants and predict whether they are likely to repay a loan.  \\ \hline
\end{tabular}
\caption{Scenario descriptions for all five AI deployment scenarios we considered in our experiments.}
\label{ref:scenario-description}
\end{table}

\begin{table}[h]
\centering
\scriptsize
\begin{tabular}{p{2.5cm}|p{12.5cm}}
\hline
 & \textbf{Stakeholders} \\ \hline
\rowcolor[HTML]{EFEFEF} 
\textbf{Communication  compliance} 
& the employee (sender),  the employee (receiver), the employees (as a group),   the family/friends of the employee,  the manager, the company,  the HR team, the legal team,  the AI system developers, the employees who identify as racial or ethnic minorities,  the employees who identify as women, the employees who identify both as women and as racial or ethnic minorities    \\

\textbf{Content  moderation}                            
& the user writing the social media post, the users who were mentioned in the post,  the content moderators, the social media company, other social media companies, the employees of the social media company, other social media platform users, the family/friends of the user writing the social media post, the AI system developers, the online community, the users who identify as racial or ethnic minorities, the users who identify as women, the users who identify both as women and as racial or ethnic minorities \\

\rowcolor[HTML]{EFEFEF} 
\textbf{Disease diagnosis} 
& the patient, other patients, future patients, the doctor, nurses, other doctors, the medical community, other patients suffering from the same disease, the clinc, other clinics, the health insurance companies, the family/friends of the patient, the AI system developers, the patients who identify as racial or ethnic minorities, the patients who identify as women, the patients who identify both as women and as racial or ethnic minorities \\

\textbf{Hiring}                                                                             
& the applicant, other applicants, future applicants, the hiring manager, the HR team,  the company, the AI system developers, the family/friends of the applicant, the applicants who identify as racial or ethnic minorities, the applicants who identify as women, the applicants who identify both as women and as racial or ethnic minorities  \\

\rowcolor[HTML]{EFEFEF} 
\textbf{Loan application}
& the applicant, other applicants, the employees of the bank, the bank, other banks, the AI system developers, the family/friends of the applicant, society, the applicants who identify as racial or ethnic minorities, the applicants who identify as women, the applicants who identify both as women and as racial or ethnic minorities  \\ \hline
\end{tabular}
\caption{The sets of stakeholders we used in our experiments for each of the five AI deployment scenarios.}
\label{table:stakeholders}
\end{table}

\newpage
%\subsection{Crowdsourcing task (include figure)}
\subsection{Crowdsourcing task}

\para{Demographic/Background questions} that accompanied our crowdsourcing task include:
\begin{compactenum}[1)]
    \item Have you experienced discrimination on the basis of your race, ethnicity, gender, nationality, sexual orientation, ability or religious beliefs? 
    \item Have you experienced any adverse impacts from any AI or computational systems you have had to use in the past? 
    \item Do you have any experience with AI systems like the one in scenario above? 
    \item How familiar are you with how AI systems like the one in the scenario above work?
    \item What is your age range? 
    \item What is the color of the sky? (Attention check question)
    \item Please let us know if you have any comments/feedback for improving this task.
\end{compactenum}
%GPT-3 prompting

\newpage
\subsection{The taxonomy of harms}
\label{appendix:harms_taxonomy}

\begin{table}[h]
\scriptsize
\centering
\begin{tabular}{@{}p{2.5cm}|p{13cm}@{}}

%\toprule
\textbf{Code}  & \textbf{Description:} If the generated example  \\ \midrule
\rowcolor[HTML]{EFEFEF}
\multicolumn{2}{p{15.5cm}}{\textbf{Quality of service harms:} when either the AI system or a service the AI system is used for doesn't work equally well for different individuals or groups or doesn't work as intended.} \\ \midrule

inadequate service                                                          &  describes stakeholders receiving inaccurate or invalid assessments that lead to low or degrading quality services due to the output of an AI system  \\
unfair treatment                                                            &  describes a stakeholder is feeling or is being treated unfairly, including due to race, ethnicity, age, gender, or other demographic or group attribute \\
%\rowcolor[HTML]{EFEFEF} 
lost content                                                                &  describes a stakeholder losing content or work due to the output of an AI system \\
unaware of harmful behaviour                                                                    &  describes how the output of an AI system can lead to a stakeholder becoming or remaining unaware of their own harmful behavior \\
%\rowcolor[HTML]{EFEFEF} 
underperforming AI                                                          &  describes the AI system not performing according to its intended usage or according to expectations \\\midrule

\rowcolor[HTML]{EFEFEF}
\multicolumn{2}{p{15.5cm}}{\textbf{Representational harms:} when the AI system's output contains stereotypical/demeaning/erasing descriptions, depictions, or representations of people, cultures, or society} \\ \midrule

%\rowcolor[HTML]{EFEFEF} 
reinforcing stereotypes                                                   & describes how the AI system output perpetuates, amplifies,  or reinforces existing stereotypes \\
target of toxic language         &  describes a stakeholder being explicitly or implicitly targeted by toxic language, or being exposed to toxic language or exposing others to toxic language \\ \midrule

\rowcolor[HTML]{EFEFEF}
\multicolumn{2}{l}{\textbf{Well-being harms:} When a stakeholder's physical, mental or general well-being is or might be affected by the output of an AI system}\\ \midrule
%\rowcolor[HTML]{EFEFEF} 
physical health                                                             &  describes a stakeholder's health condition deteriorating or being affected as a result of the output of the AI system \\
mental health                                                               &  describes a stakeholder experiencing some form of emotional or psychological distress distress or an increase in their emotional or psychological distress, including worries, feeling shame, anxiety or stress \\
%\rowcolor[HTML]{EFEFEF} 
quality of life                                                             &  describes a stakeholder's (typically an individual, group or community) needs not being fulfilled as a result of the output of the AI system   \\
loss of motivation                                                          &  describes a stakeholder feeling discouraged or losing motivation because of the output of the AI system \\
%\rowcolor[HTML]{EFEFEF} 
self-doubt                                                                  &  describes stakeholders reevaluating their abilities or worth, or becoming reluctant or more careful with their behavior due to the AI system's output   \\
safety                                                                      &  describes a stakeholder feeling unsafe or having their safety being threatened as a result of the AI system's output  \\
%\rowcolor[HTML]{EFEFEF} 
commiseration                                                      &  describes how the AI system's output can lead a stakeholder feeling sorrow, regret or empathy for other stakeholders\\ \midrule

\rowcolor[HTML]{EFEFEF}
\multicolumn{2}{l}{\textbf{Legal and reputational harms}: When the legal position or reputation of a stakeholder is or might be affected by the output of an AI system} \\ \midrule
%\rowcolor[HTML]{EFEFEF} 
distrust and reputational damage                                            &  describes a stakeholder reputation being damaged, or to a stakeholder distrusting experts, institutions or other stakeholders \\
backlash                                                                    &  describes a stakeholders experiencing repercussion or backlash from other people \\
%\rowcolor[HTML]{EFEFEF} 
legal repercussions                                                         &  describes a stakeholder experiencing or being likely to experience legal repercussions \\
scapegoat                                                                   &  describes stakeholders being blamed or held responsible for the AI system failure or for the mistakes of others  \\
%\rowcolor[HTML]{EFEFEF} 
liability                                                                   &  describes stakeholders being held responsible for their own actions \\ \midrule

\rowcolor[HTML]{EFEFEF}
\multicolumn{2}{p{15.5cm}}{\textbf{Social and societal harms:} When the AI system's output affects or might affect relationships, social institutions, communities, or can erode social and democratic structures} \\ \midrule
public health                                                               &  describes the possibility of a disease spreading on a larger scale as a result of the AI systems' failure \\
%\rowcolor[HTML]{EFEFEF} 
eroding relationships                                                       &  describes how the output of the AI system leads to miscommunication between stakeholders, a stakeholder feeling misunderstood or misunderstanding others, stakeholders' relationship being affected, or stakeholders losing contact with a friend or acquaintance \\
bad actors                                                                  &  notes or describes a situation where one stakeholder takes advantage of other stakeholders or of their situation  \\
%\rowcolor[HTML]{EFEFEF} 
social issues                                                               &  describes how the output of the AI system can lead to crime or other social problems \\
toxic environment                                                           &  describes how the AI system's output leads or can lead to a toxic environment being created in the workplace or in a community       \\ \midrule

\rowcolor[HTML]{EFEFEF}
\multicolumn{2}{p{15.5cm}@{}}{\textbf{Loss of rights or agency:} When the AI system's output leads or might lead to a loss of agency or control over aspects of someone's life, loss of privacy, or loss of other human rights}\\ \midrule
%\rowcolor[HTML]{EFEFEF} 
loss of agency                                                              &  describes how an AI system can lead to a stakeholder autonomy and agency being at risk, or to the stakeholder not feeling in control of making their own decisions  \\
loss of privacy                                                             &  describes how the AI system's output leads to a stakeholder risking to or losing their privacy \\
%\rowcolor[HTML]{EFEFEF} 
loss of rights                                                              &  describes how the AI system's output leads to a stakeholder risking to or losing various rights \\ \midrule

\rowcolor[HTML]{EFEFEF}
\multicolumn{2}{p{15.5cm}}{\textbf{Allocational harms:} When the AI system's output affects the allocation of resources or opportunities relating to finance, education, employment, healthcare, housing, insurance, or social welfare} \\ \midrule
creating busywork                                                           &  describes or can lead to the stakeholder having to work for longer (i.e., spending more time) or on more tasks than necessary \\
%\rowcolor[HTML]{EFEFEF} 
economic strain                                                             &  describes or can lead to stakeholders (e.g., company, applicant, family) experiencing economic strain, losing money/customers/business, or dealing with raising costs  \\
job security                                                                &  describes the possibility that a stakeholder might lose their job as a result of the AI system's output \\
%\rowcolor[HTML]{EFEFEF} 
waste                                                                       &  describes how the AI system leads or can lead to a stakeholder feeling that they have wasted time or other resources   \\
work satisfaction/fit                                                       &  describes how the AI system's output can lead to stakeholders not getting a suitable position or not being satisfied with their job   \\
%\rowcolor[HTML]{EFEFEF} 
productivity loss                                                           &  describes how the AI system's output can lead to productivity loss in the workplace, including due to unqualified employees, lack of diversity, a stakeholder underperforming, or lower work quality  \\
opportunity loss                                                            &  describes how the AI system's output can lead to stakeholders missing out on an opportunity, including failing to hire the most qualified individuals   \\
%\rowcolor[HTML]{EFEFEF} 
 lack of access to information                                              &  describes a stakeholder not having, being denied or losing access to information  \\
banned from site                                                            &  describes stakeholders being banned from using a forum or platform     \\ \midrule

\rowcolor[HTML]{EFEFEF}
\multicolumn{2}{l}{\textbf{Other harms:} When it is unclear what the specific harm might be due to underspecification, or when it generally talks about the amplification or exacerbation of harms} \\ \midrule
%\rowcolor[HTML]{EFEFEF} 
underspecified or repeated harm                                             &  does not elaborate upon what the specific harm might be or the example is merely repeating the premise           \\
snowball effect                                                             &  describes how the AI system's output can lead to some stated or unstated harms being amplified or perpetuated \\\midrule

\rowcolor[HTML]{EFEFEF}
\multicolumn{2}{l}{\textbf{Not meaningful:} When what is described is not a harm} \\ \midrule
%\rowcolor[HTML]{EFEFEF} 
not a harm                                                                  &  is meaningful in context but it does not represent an actual harm to any of stakeholders   \\
nonsensical                                                                   &  is not clear and the reasoning doesn't make sense in context \\ \bottomrule

\end{tabular}
\caption{Harms taxonomy, including top and second level harm categories as well as corresponding working definitions}
\end{table}

\newpage
\subsection{Examples of harm for each category in our taxonomy}

\begin{table}[h!]
\scriptsize
\centering
\def\arraystretch{0.85}
\begin{tabular}{@{}p{3cm}|p{13cm}@{}}

%\toprule
\textbf{Code}  & \textbf{Example harm:} (Scenario -- Problematic AI Behavior -- Stakeholder)   \\ \midrule
\rowcolor[HTML]{EFEFEF}
\multicolumn{2}{p{15.5cm}}{\textbf{Quality of service harms:} when either the AI system or a service the AI system is used for doesn't work equally well for different individuals or groups or doesn't work as intended.} \\ \midrule

inadequate service                                                          & (Disease Diag. -- FP -- the patient)  ``the clinic will accidentally delay your treatment, resulting in prolongation of patient's suffering or even death''  \\
unfair treatment                                                            &  (Loan Appl. -- FN -- women applicants) ``society might view you as less capable than men and as less worthy of financial incentives'' \\
%\rowcolor[HTML]{EFEFEF} 
lost content                                                                &  (Comm. compl. -- FP -- the employee (sender)) ``your email might be suppressed and people you email will then never respond back'' \\
unaware of harmful behaviour                                                                    &  (Content mod. -- FN -- women users) ``you may use toxic language unintentionally and the system will not censor it, potentially leading to social ostracization.'' \\
%\rowcolor[HTML]{EFEFEF} 
underperforming AI                                                          &  (Loan Appl. -- FN -- women \& racial minority appl.) ``you might have been given the loan if the system had accurately assessed your qualifications.'' \\\midrule

\rowcolor[HTML]{EFEFEF}
\multicolumn{2}{p{15.5cm}}{\textbf{Representational harms:} when the AI system's output contains stereotypical/demeaning/erasing descriptions, depictions, or representations of people, cultures, or society} \\ \midrule

%\rowcolor[HTML]{EFEFEF} 
reinforcing stereotypes                                                   & (Disease diag. -- FN -- racial \& ethnic minority patients) ``my group may be profiled as not likely to have it so tests are not ordered'' \\
target of toxic language         &  (Content mod. -- FN -- racial \& ethnic minority users) ``I could read offensive language about things that offend me'' \\ \midrule

\rowcolor[HTML]{EFEFEF}
\multicolumn{2}{l}{\textbf{Well-being harms:} When a stakeholder's physical, mental or general well-being is or might be affected by the output of an AI system}\\ \midrule
%\rowcolor[HTML]{EFEFEF} 
physical health                                                             &  (Disease diag. -- FN -- women \& racial minority patients) ``then I might perish without any recourse to a medical solution.'' \\
mental health                                                               &  (Hiring -- FN -- family/friends of the applicant) ``you might feel frustrated knowing that your loved one is not qualified or that they are being treated unjustly'' \\
%\rowcolor[HTML]{EFEFEF} 
quality of life                                                             &  (Loan appl. -- FN -- the applicant) ``I need the money for something and then I wouldn't be able to pay my bills or I may miss out on a deal for a car.''  \\
loss of motivation                                                          &  (Content moder. -- FP -- the online community) ``users might miss a discussion. The original poster may be discouraged from additional posts, other users may feel like the community is toxic.'' \\
%\rowcolor[HTML]{EFEFEF} 
self-doubt                                                                  &  (Content moder. -- FN -- the content moderators) ``you might feel you aren't doing your job correctly''   \\
safety                                                                      &  (Comm. compliance -- FN -- the employee (receiver)) ``you might feel harassed by the person sending the emails''  \\
%\rowcolor[HTML]{EFEFEF} 
commiseration                                                     &  (Disease diag. -- FP -- family/friends of the patient) ``I could have a friend or family member taking treatment that could hurt them because they don't need that treatment. I don't want to watch my family/friend suffer.''\\ \midrule

\rowcolor[HTML]{EFEFEF}
\multicolumn{2}{l}{\textbf{Legal and reputational harms}: When the legal position or reputation of a stakeholder is or might be affected by the output of an AI system} \\ \midrule
%\rowcolor[HTML]{EFEFEF} 
distrust and reputational damage                                            &  (Disease diag. -- FN -- the medical community) ``professionals from other communities will be motivated to cast doubt on the effectiveness of your system''\\
backlash                                                                    &  (Content moder. -- FP -- the employees of the social media company) ``I will be perceived as part of the machinery of a toxic and possibly racist/sexist/politically biased/etc. corporation and will be the target of criticism both on and offline.'' \\
%\rowcolor[HTML]{EFEFEF} 
legal repercussions                                                         &  (Comm. compl. -- FN -- the legal team) ``I will end up having to defend the company from lawsuits that will be difficult if not impossible to win.'' \\
scapegoat                                                                   &  (Loan appl. -- FP -- family/friends of the applicant) ``you might have to support your family member/friend who may not be able to repay the loan''  \\
%\rowcolor[HTML]{EFEFEF} 
liability                                                                   &  (Hiring -- FP -- the hiring manager) ``you could end up having to take responsibility of mistakes made by unqualified workers.''\\ \midrule

\rowcolor[HTML]{EFEFEF}
\multicolumn{2}{p{15.5cm}}{\textbf{Social and societal harms:} When the AI system's output affects or might affect relationships, social institutions, communities, or can erode social and democratic structures} \\ \midrule
public health                                                               &  (Disease diag. -- FN -- the medical community) ``the patient may spread the disease'' \\
%\rowcolor[HTML]{EFEFEF} 
eroding relationships                                                       &  (Content moder. -- FN -- family/friends of the user writing the social media post) ``either you may stop providing support to your family member/friend or your family/friend may be hurt when others point out the toxic language post'' \\
bad actors                                                                  &  (Disease diag. -- FP -- other employees) ``the clinic may over-diagnose the disease just to obtain more treatment fees''  \\
%\rowcolor[HTML]{EFEFEF} 
social issues                                                               &  (Content moder. -- FN -- the online community) ``The toxic language may have long term impacts on the behavioral/mental health of the online community.'' \\
toxic environment                                                           &  (Comm. compl. -- FP -- women employees) ``your manager might feel offended by what you send which will create undesirable tension in the workplace''      \\ \midrule

\rowcolor[HTML]{EFEFEF}
\multicolumn{2}{p{15.5cm}@{}}{\textbf{Loss of rights or agency:} When the AI system's output leads or might lead to a loss of agency or control over aspects of someone's life, loss of privacy, or loss of other human rights}\\ \midrule
%\rowcolor[HTML]{EFEFEF} 
loss of agency                                                              &  (Disease diag. -- FN -- the patient) ``you will be prescribed the wrong treatment and will feel like you are being treated as a guinnea pig for AI applications''  \\
loss of privacy                                                             &  (Comm compl. -- FP -- the employee(receiver)) ``I would be concerned about a lack of personal review that could potentially lead to consequences. Also, I would feel under surveillance and misunderstood by an unfeeling machine.'' \\
%\rowcolor[HTML]{EFEFEF} 
loss of rights                                                              &  (Content moder. -- FP -- other social media platform users)``freedom of speech was violated'' \\ \midrule

\rowcolor[HTML]{EFEFEF}
\multicolumn{2}{p{15.5cm}}{\textbf{Allocational harms:} When the AI system's output affects the allocation of resources or opportunities relating to finance, education, employment, healthcare, housing, insurance, or social welfare} \\ \midrule
creating busywork                                                           &  (Hiring -- FP -- the manager) ``you may feel that your employee is writing toxic emails to all his/her colleagues and you`ll feel helpless in resolving the issue and angry that you need to spend more time in his/her case instead of handling other employees'' \\
%\rowcolor[HTML]{EFEFEF} 
economic strain                                                             & (Content moder. -- FP -- the social media company) ``people will complain about your company, you may lose customers'' \\
job security                                                                &  (Loan appl. -- FN -- the employees of the bank) ``your employer might think that you were at fault in the faulty decision, and you might be punished or fired'' \\
%\rowcolor[HTML]{EFEFEF} 
waste                                                                       &  (Hiring -- FP -- women \& racial minority applicants) ``you could have your time wasted by coming to an interview you are not even qualified or even hired and then overwhelmed by a job you cannot do.''  \\
work satisfaction/fit                                                       &  (Hiring -- FP -- the hiring manager) ``I could be hiring someone who isn't properly adequate with knowledge and that person could ruin a program or create a security risk.''  \\
%\rowcolor[HTML]{EFEFEF} 
productivity loss                                                           &  (Comm. compl. -- FP -- the manager) ``time may be wasted by the false toxicity alarms, which could delay important work''  \\
opportunity loss                                                            &  (Comm. compl. -- FN -- the HR team) ``it will become nearly impossible to hire qualified candidates due to the negative perception of the workplace.''   \\
%\rowcolor[HTML]{EFEFEF} 
 lack of access to information                                              &  (Content moder. -- FP -- the online community) ``important messages might never get posted. Open communication in the online forum will be hindered and some members may be unfairly banned from the community.''  \\
banned from site                                                            &  (Content moder. -- FP -- the user writing the social media post) ``your post may be deemed as spam, you might be suspended from using the platform and other users might stop trusting you''   \\ \midrule

\rowcolor[HTML]{EFEFEF}
\multicolumn{2}{l}{\textbf{Other harms:} When it is unclear what the specific harm might be due to underspecification, or when it generally talks about the amplification or exacerbation of harms} \\ \midrule
%\rowcolor[HTML]{EFEFEF} 
underspecified or repeated harm                                             &  (Loan appl. -- FP -- society) ``it is bad things to the loan applicants''           \\
snowball effect                                                             &  (Comm. compl. -- FN -- racial \& ethnic minority employees) ``your emails might be perceived by the manager to contain toxic language; this could taint your reputation and trigger a series of events that could result in punishment'' \\\midrule

\rowcolor[HTML]{EFEFEF}
\multicolumn{2}{l}{\textbf{Not meaningful:} When what is described is not a harm} \\ \midrule
%\rowcolor[HTML]{EFEFEF} 
not a harm                                                                  &  (Hiring -- FN -- family/friends of the applicant) ``your friend/family member might take loans from other sources other than banks''  \\
nonsensical                                                                   &  (Loan appl. -- FP -- society) ``many loan seekers would take out loans from your bank, making the government think that lending the money to your bank is not financially viable'' \\ \bottomrule

\end{tabular}
\caption{Examples of harms generated with AHA!}
\end{table}

\newpage

%\newpage
\subsection{Interview materials}
\label{appendi:interface}

\begin{figure}[h]
\centering
%\begin{subfigure}[b]{\linewidth}
\includegraphics[width=\linewidth]{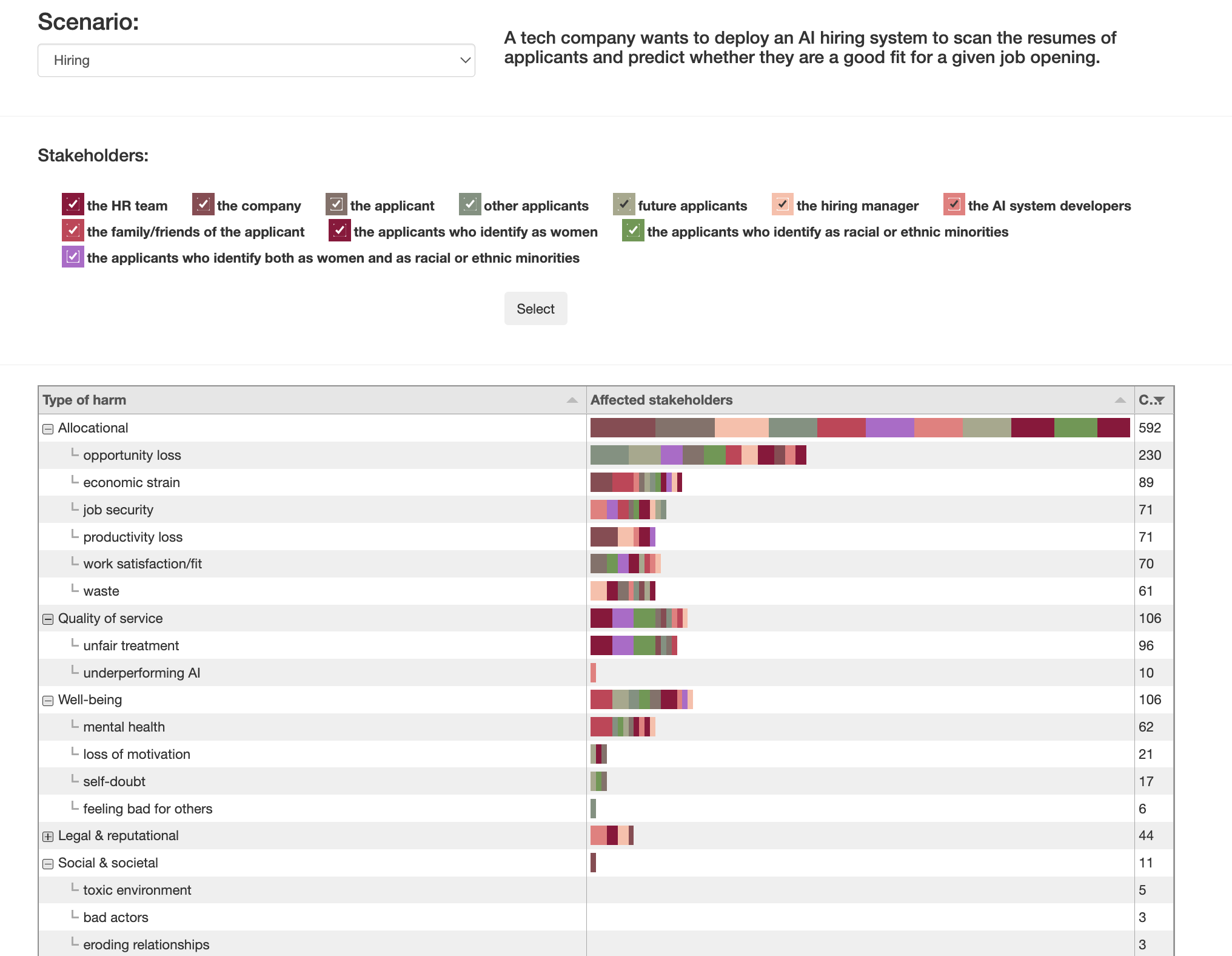}
%\caption{Code-level view} \label{fig:interface}
%\end{subfigure}

%\begin{subfigure}[b]{\linewidth}
%\includegraphics[width=\linewidth]{figures/harm-view.png}
%\caption{Harm-level view. Expanding harms under \em legal repercussions \em for \em the hiring manager \em as stakeholder.} \label{fig:interface-harm-level}
%\end{subfigure}
\caption{The interactive interface used during interviews to allow participants navigate through the harm examples surfaced by \frameworkname{}. Participants could select stakeholders and filter the examples of harms accordingly.}
\label{fig:interface}
\vspace{-12pt}
\end{figure}

\end{document}